\documentclass[usletter, 12pt]{article}

\usepackage{amsmath,amsthm}
\usepackage{natbib}
\usepackage{geometry}
\usepackage{amssymb,authblk,amsfonts}
\usepackage{comment,enumerate,graphicx,hhline}

\usepackage[usenames,dvipsnames]{color, xcolor}

\usepackage{subcaption}
\usepackage[hidelinks]{hyperref}

\usepackage{amsfonts,amstext,rotating,amssymb,graphics,xspace}

\usepackage{epsfig}
\usepackage{hyperref}
\hypersetup{
	colorlinks   = true, 
	urlcolor     = blue, 
	citecolor   = blue    
}

\usepackage{mathptmx}
\usepackage{type1cm}
\usepackage{mathrsfs}
\usepackage{amsmath}
\usepackage{natbib}
\usepackage{hyperref}
\usepackage{geometry}
\usepackage{type1cm}
\usepackage{mathrsfs}
\usepackage{amssymb}
\usepackage{authblk}
\usepackage{amsfonts}
\usepackage[T2A,T1]{fontenc}
\usepackage{hhline}
\usepackage{multirow}
\usepackage{multicol}
\usepackage{color}
\usepackage{sgame}
\usepackage{mathtools}
\usepackage{caption}
\usepackage{graphicx}
\usepackage{wrapfig}
\usepackage{changepage}
\usepackage{lipsum}
\usepackage[english]{babel}
\usepackage[utf8]{inputenc}
\usepackage{comment}
\usepackage{tikz}
\usepackage{pgf}
\usetikzlibrary{positioning}
\usepackage{tabularx}
\usepackage{pbox}
\usepackage{subcaption}
\usepackage{rotating}
\usepackage{tcolorbox}
\usepackage{booktabs}
\usepackage{adjustbox}
\usepackage{appendix}

\definecolor{mypurple1}{RGB}{204,51,255}

\usetikzlibrary{arrows,backgrounds}

\tikzstyle{object}=[circle,draw=red]
\tikzstyle{agent}=[circle,draw=blue]
\tikzstyle{quantity}=[fill=white]

\def\bv{\mathbf{v}}

\newcommand{\hide}[1]{}

\hypersetup{colorlinks = true}
\hypersetup{allcolors=blue}

\newtheorem{proposition}{Proposition}
\newtheorem{lemma}{Lemma}
\newtheorem{example}{Example}

\newtheorem{hypothesis}{Hypothesis}

\usepackage{setspace}

\begin{document}
	
	\title{Making school choice lotteries transparent}
	\author{Lingbo Huang\thanks{Center for Economic Research, Shandong University, Jinan, China. Email: lingbo.huang@outlook.com.} \quad Jun Zhang\thanks{Institute for Social and Economic Research, Nanjing Audit University, Nanjing, China. Email: zhangjun404@gmail.com.}}
	
	\date{\today}
	
	\maketitle
	\begin{abstract}
		
		Lotteries are commonly employed in school choice to fairly resolve priority ties; however, current practices typically keep students uninformed about their lottery outcomes at the time of preference submission. This paper advocates for revealing lottery information to students beforehand. When preference lists are constrained in length, which is a common feature in real-world systems, such disclosure reduces uncertainty and enables students to make more informed decisions. We demonstrate the benefits of lottery revelation through two stylized models. Theoretical predictions are supported by laboratory experiments. 
	\end{abstract}
	
	\bigskip
	
	
	

	\thispagestyle{empty}
	\setcounter{page}{0}
	\newpage

	\section{Introduction}
	
	People need information to make informed decisions. We apply this logic to guide policy reforms in school choice.
	Many cities worldwide use centralized school admission mechanisms and assign school priorities to students based on broad criteria (geographic proximity, demographic characteristics, and other factors). 
	Priority ties frequently occur within these systems. Popular mechanisms, such as the Boston and the Deferred Acceptance (DA), however, require strict priorities for input. So, to guarantee fairness, those cities often resort to lotteries to resolve priority ties. Despite the crucial role that lotteries play in determining student assignments, in current practices, students are typically unaware of their lottery outcomes when submitting preferences. To our knowledge, this issue has been overlooked in the school choice literature, presumably for two reasons. First, many studies assume strict priorities from the outset, thereby eliminating the need to consider lotteries. Second, most studies assume complete preference lists and advocate for strategy-proof mechanisms, under which students have no need to acquire any information. However, in practice, preference lists are frequently constrained in length.\footnote{Additionally, forming a complete ranking of schools is itself often challenging: acquiring meaningful information about each school can be costly, time-consuming, and unevenly accessible across students. As a result, preference lists are not only administratively capped but also de facto constrained by these information-acquisition frictions. We will return to this point in the later part of this section.} According to \cite{neilson2024rise}, who reviews primary,
	secondary, and higher education markets in 149 countries, Chile is the only country that employs a centralized strategy-proof mechanism and an unlimited preference list in primary education; only a few other countries do so for secondary and higher education.  
	Under constrained preference lists, no mechanisms can be truly strategy-proof, as students must strategically choose which schools to rank, and the risk of going unmatched becomes a nonnegligible issue. In New York City (NYC), while students can rank up to 12 schools, 7\% (around 5,000 students) remained unmatched after the main round of the match during the years 2021 and 2022. In certain districts, the unmatch rate can reach as high as 18\%. 

	This paper addresses the above issue in practical school choice. We advocate for a ``reveal'' policy in which lotteries are drawn and revealed to students before their submission of preferences, contrasting it with the current ``cover'' policy, whereby students are uninformed about their lotteries before submitting preferences. To illustrate the different consequences of the two policies, consider two students $ i $ and $ j $ competing for two schools A and B, each with a single seat. Both students prefer A over B, and they are in priority ties for both schools. A single lottery is used to break the ties. Students are limited to reporting only one school. Under the cover policy, provided that both students value A sufficiently higher than B, both will report A. This is a rational choice, despite their awareness of the risk of being unmatched. 
	In contrast, under the reveal policy, students can effectively coordinate their strategies upon learning their lottery numbers: the student favored by the lottery outcome will report A and the other will report B. Consequently, both students will be matched, and \textit{ex ante}, each still has an equal chance of being admitted to A.

	The intuition for the reveal policy is simple: providing students with their lottery information enables more informed preference reporting. This paper employs stylized models and laboratory experiments to demonstrate this intuition. Notably, the reveal policy  is not only conceptual; it has been implemented in practice. During the preparation of this paper, we observe that NYC has adopted this policy since the 2022-2023 season. Initially, NYC refused to reveal lottery numbers. However, following a parent-led campaign under the New York State's Freedom of Information Law, NYC first agreed to reveal lotteries upon request after admissions and finally decided to disclose lotteries before preference submissions. Therefore, the results of this paper provide justification for NYC's recent policy change.

	To demonstrate the benefits of the reveal policy transparently, we study two variants of the canonical model of \cite{abdulkadiroglu2011resolving} (ACY hereafter) and show that the reveal policy can effectively resolve uncertainties for students and simplify their preference submission strategies by informing them of the ``right'' schools to rank. 
	In ACY's original model, students have common ordinal preferences over schools but differ in their privately known cardinal utilities. All students are in priority ties for all schools, and a single lottery determines the ranking of students for all schools. The DA mechanism determines the matching. This simple priority structure and the conflicting preferences setting together are particularly well-suited for demonstrating the role of lotteries.
	
	Our first model deviates from ACY's by restricting the number of schools that students can rank. Thus, students need to decide which schools to rank. 
	We show that a simple reveal policy, termed \textbf{Reveal policy}, in which students are informed of their own lottery numbers but not the others',\footnote{We assume that the number of students and the capacities of schools in our model are common knowledge.} is sufficient to resolve uncertainties for students. Each student must rank and is assigned to his  best achievable school, resulting in a matching equivalent to that produced by DA under complete preference lists. Consequently, every student is always matched. From both the \textit{interim} (after utilities are realized but before the lottery is drawn) and the \textit{ex-ante} (before utilities and the lottery are drawn) perspective, students have equal chances of attending each school. In contrast, under the \textbf{Cover policy}, students' strategies can be complicated. For certain realized utilities, students may concentrate their submissions on a subset of schools, resulting in wasted seats at other schools and leaving some students unmatched. From the \textit{interim} perspective, every student faces a positive probability of being unmatched, and from the \textit{ex-ante} perspective, students receive a random assignment that is first-order stochastically dominated by that obtained under the Reveal policy.

	Our second model brings the analysis closer to real-world settings, where schools often implement nontrivial priority policies. A frequently observed priority policy in the real world is neighborhood. So, we extend the first model by incorporating neighborhoods: each school prioritizes a subset of students living in its neighborhood, while still resolving ties within the same student type by a single lottery. The presence of neighborhoods complicates the reveal policy. Consider a student who, upon learning his lottery number, applies to a school where he lacks neighborhood priority. In the first model, his lottery number directly indicates his ranking relative to all others across all schools. In the second model, however, he faces uncertainty about how many students with worse lottery numbers nonetheless hold neighborhood priority for the school. In other words, his lottery number no longer provides a clear signal of his chances at each school. In practice, this issue may be solved in several ways. For instance, if the student population is large and their preference distribution is predictable, the uncertainty diminishes due to the law of large numbers. Moreover, if students’ preference distribution remains relatively stable over time, the cutoff lottery numbers for schools in admission outcomes will tend to stabilize across years.\footnote{While this argument appears to be new in the context of lottery-based school choice, it has been empirically verified in test-based admission systems (e.g., China and Turkey). The underlying theoretical insight has been provided by \cite{azevedo2016supply}, who show that a continuum market generically has a unique stable matching, which is the limit of stable matchings in large finite markets.} In such settings, students can use past cutoff data along with their lottery number to estimate their chances at different schools. 
	
	Our second model considers another approach policymakers may adopt: providing students with additional statistics about lottery outcomes. To illustrate this idea, we propose a more informative version of the reveal policy, termed \textbf{RevealMore policy}, in which policymakers publicly disclose, for each neighborhood, the number of students who receive lottery numbers weakly above each possible value.\footnote{Similar policies are used in China’s college admissions system, where students are privately informed of their exam scores while the government publishes the number of students scoring above each possible value.} These statistics allow students to infer the distribution of lottery numbers across neighborhoods. We show that the RevealMore policy can replicate the effect of the Reveal policy in the first model by effectively informing students of their achievable schools. In contrast, under the Cover policy, students with certain utilities choose to remain in their neighborhoods to minimize risk, which undermines the objective of school choice.\footnote{The experiment of \cite{calsamiglia2010constrained} provided evidence that under the Cover policy, compared to unconstrained preference lists, constrained preference lists cause students to exhibit a greater neighborhood school bias and increased segregation.} Interestingly, we observe that the RevealMore policy has been somehow adopted by NYC in the 2024-2025 season: in the application platform (MySchools), for each program, a student can see an icon next to the program indicating whether he has a low, medium, or high chance of getting accepted into the program, which is calculated based on last year’s admissions process, the student's characteristics, and his assigned lottery number.\footnote{Students will see one, two, or three orange bars next to each program. Three orange bars means that a student has a high chance of receiving an offer to this program. Two orange bars means a medium chance, and one orange bar means a low chance.}
	
	In both models, students' strategies under the reveal policies are straightforward, whereas their strategies under the cover policy are far less transparent. This is not surprising. Prior literature has shown that, even under strict priorities and complete information, students' strategies under DA with constrained preference lists are difficult to characterize  \citep{haeringer2009constrained,decerf2021manipulability}. Only recently have \cite{ali2025hedging} provided a framework for understanding the portfolio of schools students should rank when school admissions are correlated.  The portfolios students rank depend on their subjective beliefs. It is conceivable that more pessimistic students might apply less aggressively, leading to less favorable outcomes. Therefore, the reveal policy may help level the playing field by eliminating disparities arising from differences in subjective beliefs, though this potential benefit falls outside the scope of this paper.

	To test our models and quantify the differences between various policies, we conduct two laboratory experiments. The main experiment adopts a relatively simple environment in which we vary the presence of neighborhood schools and the lottery revelation policy, and students rank a single school in their preference lists. This one-school setting is extremely simple, but it is necessary to characterize students' equilibrium strategies in all treatments. We derive precise predictions regarding the expected match rates and expected student payoffs. The experimental results indicate that the Reveal policy outperform the Cover policy by yielding significantly higher match rates and student payoffs, regardless of the presence of neighborhood schools. Further, the Reveal policy significantly increases the likelihood that students with various lottery numbers adhere to equilibrium strategies. Notably, when neighborhood schools are present, the RevealMore policy enhances market efficiency and reduces deviations from equilibrium strategies, as predicted, compared to the Reveal policy. The robustness experiment implements a more complex environment with a larger market and permits students to rank up to two schools. The superiority of the reveal policy persists across different strategic environments.
	

	Overall, our results support the reveal policy and identify the key channel through which it improves the match process and outcome. To keep our analysis concise and focused, we make simplifying assumptions in models and experiments. In particular, the reveal policy in our models can eliminate all uncertainties for students. In practice, however, it is likely that the reveal policy can reduce but not entirely eliminate uncertainties. To what extent the policy can help students predict their achievable schools in practical markets is an empirical question. Since real-world markets are much larger and more complex than any laboratory experimental setting, this question has to be answered using field data. This represents an exciting direction for future research. Using a crowdsourced survey, \cite{marian2023algorithmic} demonstrates the positive impact of NYC's reveal policy on improving the match outcomes for survey participants. Her finding provides an empirical support for the insight in our paper.
	
	Our analysis assumes that students face constrained preference lists, which holds widely in the real world. But since list length is itself a policy choice, this raises an immediate question: if policymakers seek to assist students, why not simply extend list lengths? More generally, how should we compare the effect of revealing lottery information with the effect of lengthening lists? The second question is again empirical, while the first connects to a deeper issue in school choice. Models since the seminal work of \cite{abdulkadiroglu2003school}, including ours, typically assume that students have complete preferences before the match, but in reality students need information to form preferences and, under list constraints, must decide which schools to rank. NYC’s recent reform sheds light on this point. While NYC imposes the 12-school cap on preference list, many students already struggle to find 12 choices to list. In recent years, about 38\% of high school applicants submitted full 12-school lists. For the remaining 62\%, submitting shorter lists does not indicate that the length constraint was not binding for them; it reflects the difficulty in finding suitable schools to rank, in the absence of information. As of the 2024–2025 cycle, NYC has removed the 12-school cap on high school applications, allowing students to rank an unlimited number of schools. This change does not replace the reveal policy adopted in 2022–2023. Rather, NYC has increased the information provided to students (the icon described above) to help them evaluate admission chances at each school. 
	This practice suggests that list length is not the only issue in practice. Providing information, which includes lottery results  in the NYC case, remains essential for students to make informed decisions, even when lists are unconstrained.
	
	This paper contributes to the growing literature on information provision in school choice. We discuss this literature below. Here, we highlight recent findings from empirical studies. Although Chile allows students to rank an unlimited number of schools, \cite{arteaga2022smart} find that students on average rank only three schools, and this is because students lack information about which schools are in their choice set. Students typically overestimate their admission chances, and a significant portion of students face an over 50\% probability of non-placement. In intervention experiments, students who
	initially submit short lists add one or more schools to their lists after being warned of the risk.  Similarly, \cite{ainsworth2023households} and \cite{ajayi2025information}, using data from Romanian and Ghana, respectively, find that students often make school choice decisions with inadequate information, and, if appropriate information is provided, they respond sensitively and make better decisions. These findings suggest that researchers should not only work on the design of algorithms that produce matchings but also explore the design of information provision mechanisms. In our setting, since lottery results are free information on the platform, it could be provided at no cost to help students better estimate their choice set and secure better placements.

	\paragraph{Related literature} 
	To our knowledge, we are the first to formalize the reveal policy suggestion for lottery-based school choice. Importantly, it is not merely an idea but has already become practice in NYC, one of the world's largest school choice systems. Previous studies that consider lotteries fall mainly into three categories: (1) examining how tie-breaking affects the efficiency of stable matching mechanisms; (2) reevaluating manipulable mechanisms by incorporating the uncertainty created by tie-breaking; (3) comparing different tie-breaking rules. All these studies take the prevailing practice of withholding lottery information from students as given.
	
	Specifically, \cite{erdil2008s} and \cite{abdulkadirouglu2009strategy} show that the stability constraints created by exogenous tie-breaking can result in efficiency loss in the outcome of DA, and they propose algorithms to eliminate the efficiency loss. However, it is impossible to improve on the DA outcome without losing its strategy-proofness property.
	
	\cite{abdulkadiroglu2011resolving} compare the Boston mechanism (BM) and DA under the cover policy with unconstrained preference lists. Their main message is that students' strategic behavior under BM, in response to the uncertainties associated with tie-breaking, can result in higher welfare than DA. Our models build on ACY, but our insights differ significantly. We compare two revelation policies under DA with constrained preference lists. The benefit of the reveal policy comes from the simplified strategies of students after learning their lottery results, whereas the welfare superiority of BM in ACY's result relies on non-truth-telling equilibrium, which, as suggested by the experiments of \cite{featherstone2016boston}, might not be achievable by real-world students.
	
	There is a body of literature that compares single lotteries versus multiple lotteries (i.e., different schools may use different lotteries) for tie-breaking (e.g., \citealp{abdulkadirouglu2009strategy,pathak2011lotteries,ashlagi2020matters,arnosti2023lottery}).  While our paper operates under the assumption of a single lottery, the main insight carries over to multiple lotteries. However, under multiple lotteries, students must navigate more complex information, which leads us to believe that a single lottery is superior in this regard. Fortunately, single lotteries are more frequently employed in practice.

	The reveal policy we study parallels the reform of preference-submission timing in China's college admission system. The reform moves preference submission from before students learn their college-entrance exam scores to after scores are released. It is accompanied by a shift from sequential mechanisms (variants of BM) to parallel mechanisms (variants of DA), as studied by \cite{chen2017chinese}. These changes are widely viewed as successful in reducing uncertainty for students and improve match quality. \cite{lien2016preference,lien2017ex}, however, provide a different perspective on this reform. Assuming that the true abilities of students are not perfectly signaled by their exam scores, the authors highlight the superiority of BM with pre-exam preference submission. They argue that, when students submit their
	preferences before the exam, those students with higher abilities are more willing to apply for better schools since they have higher expected scores, while those with lower expected scores are less willing to do so. This produces a more ex-ante fair matching. Moreover, students having higher preference intensities for good schools are more willing to apply for them than those with lower preference intensities, which leads to a more ex-ante efficient matching. This latter point resembles ACY's insight.
	
	In school choice, lotteries play a role similar to exam scores, except that they are randomly generated and independent of students' effort. The reform in China gives us confidence that a reveal policy in school choice can similarly help students. NYC's adoption of this policy further suggests that policymakers have recognized its benefits. While \cite{lien2016preference,lien2017ex} provide reflections on the reform in China, their arguments do not apply to school choice, since there is no concern about students' abilities here, and since lottery numbers are random, the sorting channel central to their arguments is absent. 
	Beyond conveying different messages, the models and experiments in \cite{lien2016preference,lien2017ex} also differ from ours. They compare BM and Serial Dictatorship in one-to-one school matching with complete preferences lists. The difficulties associated with constrained preference lists are absent in their studies.

	A growing literature studies how explicitly disclosing agents' attainable choice sets affects their behavior and welfare in matching environments. A main insight is that sequentially disclosing information or presenting mechanisms in a dynamic form can improve information efficiency and simplify agents' strategies. For example, \citet{li2017obvious} demonstrates that dynamic mechanisms presenting agents with simple, staged choice sets can mitigate strategic mistakes made by boundedly rational agents. \citet{bo2020iterative,bo2024pick} show that iterative versions of DA, which require students to apply to one school at a time, outperform the static DA in truth‐telling rates and in the frequency of stable outcomes. \citet{hakimov2023costly} find that a sequential serial dictatorship mechanism induces less wasteful information acquisition and yields higher student welfare than its direct counterpart.\footnote{Other contributions in this literature include  \citet{bade2015serial}, \cite{chen2021information,chen2022information}, \citet{noda2022strategic}, \citet{mackenzie2022menu}, \citet{pycia2023theory}, among others.}  In our setting, revealing the realization of the single tie-breaking lottery turns a complicated distribution over admissions into a simple, individualized choice set and thereby makes constrained rank-order list decisions more transparent and less risky.


	Finally, our paper is related to the emerging literature on transparency in school choice and matching mechanisms. 
	\citet{hakimov2025improving} theoretically and experimentally study policies that allow students to verify their school assignments and the allocation rules under popular mechanisms. \cite{moller2025transparent} formalizes a concept of transparency and examines conditions under which mechanisms promised by policymakers with limited commitment can be trusted or deviations can be detected.\footnote{Similar concepts include credibility by \cite{akbarpour2020credible} and auditability by \cite{grigoryan2024theory} and \cite{pycia2024ordinal}.}
	While we do not study policymakers' commitment problems per se, the reveal policy effectively works as a transparency intervention in our setting. Beyond improving efficiency and fairness, this intervention may bring an additional behavioral benefit not studied in our paper: the reduction in student anxiety. When students know that lotteries play a decisive role in their assignments yet remain unobservable, they may experience uncertainty and discomfort. Making lottery results visible can mitigate this psychological burden. Indeed, this concern appears salient in practice: in New York City, parents have organized campaigns calling for public disclosure of lottery results.

	\section{Two theoretical models}\label{section:model}

	\subsection{Common priority model}
	
	A number $ n $ of students $ I=\{i_1,\cdots,i_n\} $ seeks admission to a number $ m $ of schools $ S=\{s_1,\cdots,s_m\} $. Each school $ s $ has capacity $ q_s\in \mathbb{N} $, with $ \sum_{s\in S} q_s=|I| $; that is, school seats are exactly enough to admit all students. Each student $ i $ has a utility vector $ \bv^i=(v^i_1,v^i_2,\ldots,v^i_m) $ where $ v^i_k\in [0,1] $ denotes $i$'s utility of attending school $ s_k $. Each $\bv^i$ is independently drawn from a finite set $ \mathcal{V}=\{(v_k)_{k=1}^m\in [0,1]^m:v_1>v_2>\cdots>v_m \} $ with some probability $ f(\bv^i) \in (0,1)$. Therefore, students have identical ordinal preferences: $ s_1 \succ s_2 \succ s_3 \ldots \succ s_m $. Students know their own cardinal utilities but not the others', except for the underlying probability distribution $f$.  All students are in priority ties for all schools. Priority ties are broken according to a single lottery, which is represented by a one-to-one mapping $ \pi: I\rightarrow \{1,2,\ldots,n\} $. The lottery is drawn uniformly at random. After a lottery $ \pi $ is drawn, if $ \pi(i)<\pi(j) $, it means that $ i $ is ranked above $ j $ in the lottery.  A matching is a function $ \mu: I \rightarrow S\cup \{\emptyset\} $  such that, for each school $ s $, $ |\mu^{-1}(s)|\le q_s $. For any student $ i $, if $ \mu(i)=\emptyset $, it means that $ i $ is unmatched. 
	
	So far, we have presented the setup of ACY's model. Our model deviates from theirs in that we consider constrained preference lists. Specifically, we denote by $ \ell $ the length of the rank-order list (ROL) and assume that $  \ell <m $. So, it is infeasible for students to rank all schools. We denote by $ \text{DA}^\ell $ the student-proposing DA mechanism with the constrained ROL.\footnote{The $ \text{DA}^\ell $ procedure proceeds as follows: in each step, each unassigned student applies to the best school in his reported preferences that has not rejected him, and each school tentatively admits the highest-ranked applicants among those admitted in the previous step and new applicants in this step.}

	Our purpose is to compare two policies regarding the disclosure of lottery. Under the \textbf{Cover policy}, which is prevalent in real life, students remain uninformed about their lottery outcomes before submitting ROLs. This uncertainty could complicate students' strategies. 
	
	In contrast, we advocate for the \textbf{Reveal policy}, in which students are informed of their lottery outcomes before submitting ROLs. Figure \ref{fig:timeline} shows the timeline of the school choice game under the Reveal policy: students' utilities are first realized; the lottery is then drawn and revealed to the students; students then submit their ROLs; finally, the matching outcome is found by $ \text{DA}^\ell $.
	
	In general, under the Reveal policy, policymakers may reveal various information about the lottery to reduce uncertainties for students. For instance, each student may be informed of his own lottery outcome and useful statistics about the others' lottery outcomes. For the simple model in this subsection, we assume that policymakers only reveal to each student $ i $ his own lottery number $ \pi(i) $. This is sufficient to eliminate all uncertainties for students. 
	
	Our analysis compares the two policies at different stages of the game. \textbf{Ex-ante} refers to the timing before students' utilities are realized. \textbf{Interim} refers to the timing after students' utilities are realized but before the lottery is drawn. For individual students, \textbf{interim} also denotes the timing when they only know their own utilities (but do not know the others' and the lottery outcome). \textbf{Ex-post} refers to the timing after the lottery has been drawn and students have submitted their ROLs.

	\begin{figure}[!htb]
		\centering
		\footnotesize
		\begin{tikzpicture}[x=3cm]
			\draw[black,->,thick,>=latex,line cap=rect]
			(-.5,0) -- (4.5,0);
			\foreach \Xc in {.25,1.5,2.75,4}
			{
				\draw[black,thick] 
				(\Xc,0) -- ++(0,5pt);
			}   
			\node[above,align=left,anchor=north,inner xsep=0pt] 
			at (-0.2,.6) {Ex-ante};

			\node[below,align=left,anchor=north,inner xsep=0pt] 
			at (0.25,-.2) {Utilities \\ realized};  
			

			\node[above,align=left,anchor=north,inner xsep=0pt] 
			at (1,.6) {Interim}; 
			
			\node[below,align=left,anchor=north,inner xsep=0pt] 
			at (1.6,-.2) 
			{Lottery drawn and \\  revealed to students};
			
			\node[below,align=left,anchor=north,inner xsep=0pt] 
			at (2.9,-.2) 
			{ROLs \\ submitted};
			
			
			\node[above,align=left,anchor=north,inner xsep=0pt] 
			at (3.5,.6) {Ex-post};
			
			\node[below,align=left,anchor=north,inner xsep=0pt] 
			at (4,-.2) 
			{Matching\\ generated};
		\end{tikzpicture}
		
		\caption{Timeline of the Reveal policy}\label{fig:timeline}
	\end{figure}

	We first show that students' strategies become straightforward under the Reveal policy. By observing lottery numbers, students can infer their best achievable schools and essentially rank those schools highest in their ROLs. Specifically, there exists a unique Bayesian Nash equilibrium (BNE) in which the top $ q_{s_1} $ students in the lottery rank $s_1$ highest in their ROLs and be admitted to $s_1$; the next $ q_{s_2} $ students in the lottery rank $ s_2 $ highest (or below $ s_1 $) in their ROLs and be admitted to $ s_2 $; and so on.
	
	\begin{lemma}
		In the common priority model, under the Reveal policy, for any realized utilities and any drawn lottery $ \pi $, there exists a unique BNE in which, for every student $ i $:
		\begin{itemize}
			\item If $ \pi(i)\in [1,q_{s_1}] $, then $ i $ is admitted to $ s_1 $ by ranking $ s_1 $ highest;
			
			\item If $ \pi(i)\in [ \sum_{y=1}^{k-1}q_{s_y}+1, \sum_{y=1}^{k}q_{s_y}] $ for some $ 2\le k \le m$, then $ i $ is admitted to $ s_k $ by ranking $ s_k $ essentially highest.\footnote{Student $i$ ranks $s_k$ essentially highest if he ranks $s_k$ highest or only below some schools in $ \{s_1,s_2,\ldots,s_{k-1}\} $.}    	
		\end{itemize}
	\end{lemma}
	
	\begin{proof}
		Under the Reveal policy, if a student observes that his lottery number is among the top $q_{s_1}$, he must rank $s_1$ highest in his ROL and be admitted to $s_1$. Given this, if a student observes that his lottery number is among the next $q_{s_2}$ below the top $q_{s_1}$, he knows that the top $q_{s_1}$ students must be admitted to $s_1$ and, therefore, $s_2$ is his best achievable school. So, he must rank $s_2$ highest (or only below $s_1$) in his ROL and be admitted to $s_2$. Similarly, the next $q_{s_3}$ students in the lottery must essentially rank $s_3$ highest and be admitted to $s_3$. The result holds inductively for all other students. 
	\end{proof}
	
	Therefore, revealing lottery numbers resolves uncertainties for students and coordinates their strategies. From an ex-post view, given the same lottery, $ \text{DA}^\ell $ under the Reveal policy finds the same matching as the unconstrained DA does, where students report complete and true preferences. So, the Reveal policy achieves the same effect as allowing students to report complete preferences. 
	
	From an interim view, since the lottery is drawn uniformly at random, students have equal probabilities of receiving each lottery number. Thus, they have equal probabilities of attending each school. This result also holds from an ex-ante view. Thus, we present the following result without a proof.
	
	\begin{proposition}\label{prop:reveal}
		In the common priority model, under the Reveal policy:\\
		(1) From an ex-post view, the outcome of $\text{DA} ^\ell $ coincides with the dominant strategy outcome of DA with unconstrained ROL. In particular, no students are unmatched.\\
		(2) From an interim (and ex-ante) view, students have equal probabilities of attending each school, which is equivalent to a random assignment $ (\frac{q_s}{n}\cdot s)_{s\in S} $. 
	\end{proposition}
	
	Next, we analyze the Cover policy, where students cannot base their strategies on their lottery numbers. Let $ \mathcal{L}^\ell $ denote the set of all possible ROLs that rank no more than $ \ell $ schools. A (mixed) strategy is a mapping $ \sigma: \mathcal{V}\rightarrow \Delta(\mathcal{L}^\ell) $, where $ \Delta(\mathcal{L}^\ell) $ denotes the set of probability distributions on $\mathcal{L}^\ell$. For any utility vector $ \bv^i $, $ \sigma(\bv^i) $ is a strategy in which $i$ reports $ L\in \mathcal{L}^\ell $ with probability $ \sigma(\bv^i)(L) $.
	
	We focus on symmetric BNE in which students with identical utilities play the same strategies. Let $ \sigma^* $ denote such an equilibrium. Then, for any realized utilities $ \bv=(\bv^i)_{i\in I} $, each student $ i $ plays the strategy $ \sigma^*(\bv^i) $. Although it is generally difficult to characterize $ \sigma^* $, we know that, for any $ L\in \mathcal{L}^\ell  $ such that $ \sigma^*(\bv^i)(L)>0 $, at least $ m-\ell $ schools are not listed in $ L $. Therefore, if all students have equal cardinal utilities $\bv^i$, which occurs with positive probability, the event that at least $ m-\ell $ schools are not reported by any student happens with the probability $\sum_{L\in \mathcal{L}^\ell } [\sigma^*(\bv^i)(L)]^n$. In this event, the capacity of those schools is wasted, and at least an equal number of students are unmatched.
	
	More generally, when students have different cardinal utilities, as long as some school $ s $ is not reported by any student, at least $ q_{s} $ students are unmatched. From an interim view, every student believes that he has a positive probability of being unmatched, since there is a positive probability that the others have identical utilities to his.  From an ex-ante view, because students are symmetric, they must have equal probabilities of attending each school and equal probabilities of being unmatched. So, the equivalent random assignment must be first-order stochastically dominated by the random assignment $ (\frac{q_s}{n}\cdot s)_{s\in S} $ under the Reveal policy. Thus, we present the following result without a proof.

	\begin{proposition}\label{prop:cover}
		In the common priority model, under the Cover policy:\\ 
		(1) Ex-post, with a positive probability, at least $m-\ell$ schools' capacities are wasted and at least an equal number of students are unmatched;  \\
		(2) Interim, each student believes that he has a positive probability of being unmatched; \\
		(3) Ex-ante, each student obtains an equal random assignment that is first-order stochastically dominated by the counterpart under the Reveal policy.
	\end{proposition}
	
	We provide an example in which students' equilibrium strategies are characterized. The environment in the example is used in our experiments.
	
	\begin{example}\label{example:first:model}
		There are six students $ \{1,2,3,4,5,6\} $ and three schools $ \{s_1,s_2,s_3\} $. Each school has two seats. Each student can report only one school in their ROL. There are two types of students' utilities $ \{\bv,\bv'\} $,\footnote{To be consistent with our experiments, we normalize student utilities on a scale of 1 to 100. This normalization does not affect our analysis.} which follow the distribution $ f(\bv)=2/3 $ and $ f(\bv')=1/3 $:
		\begin{table}[!htb]
			\centering
			\begin{tabular}{|c|c|c|}
				\hline
				& $ \bv $ & $ \bv' $ \\ \hline
				$ s_1 $ & $ 90 $ &  $ 70 $ \\ \hline
				$ s_2 $ & $ 40 $ &  $ 60 $ \\ \hline
				$ s_3 $ & $ 20 $ &  $ 20 $ \\ \hline			
			\end{tabular}
		\end{table}
		
		\textbf{Cover policy}: There is a unique symmetric BNE in which type-$\bv$ students report $s_1$ and type-$\bv'$ students report $s_2$. To verify that this is an equilibrium, for any student $i$, if the others follow the equilibrium strategy, then the admission probability to each school by reporting the school in the ROL is $\frac{361}{729}\approx .50$ for $s_1$, $\frac{569}{729}\approx .78 $ for $s_2$, and $1$ for $s_3$. So, every type-$\bv$ student's optimal strategy is to report $ s_1 $, and every type-$\bv'$ student's optimal strategy is to report $ s_2 $. No students report $ s_3 $. So, $s_3$ is always vacant and at least two students are not matched.
		
		From an ex-ante view, every student receives the (approximate) random assignment $ (0.33s_1,0.26s_2, 0.41\emptyset) $. The expected match rate is about $ 59\%$, and the expected utility is about $45.33$. From an interim view, type-$ \bv $ students' expected match rate is about $ 49\%$ and their expected utility is $ 44.57 $. Type-$ \bv' $ students' expected match rate is about $ 78\% $ and their expected utility is $ 46.83 $.
		

		\textbf{Reveal policy}: Students' equilibrium strategies depend only on their own lottery numbers: those receiving the best two lottery numbers $1$ and $2$ report $s_1$; those receiving lottery numbers $3$ and $4$ report $s_2$; those receiving lottery numbers $5$ and $6$ report $s_3$. All students are admitted to their reported school.
		
		From an interim view, every student obtains the random assignment $ (\frac{1}{3}s_1,\frac{1}{3}s_2,\frac{1}{3}s_3) $. The expected match rate is $100\%$ and the expected utility is $50$, regardless of students' type. This also holds from an ex-ante view.
		
		Hence, from both an interim and an ex ante view, students obtain strictly higher expected match rates and expected utilities under the Reveal policy than under the Cover policy.\footnote{But in general, the comparison between the two policies in terms of interim expected utilities is ambiguous.} 
	\end{example}

	\subsection{Neighborhood priority model}\label{theory:neighbor}

	In the previous model, we assume that all students have common priorities for all schools. This amplifies the role of lotteries but deviates from the reality. In practice,  schools often use nontrivial priority policies, among which neighborhood is one of the most frequently observed policies. To demonstrate how the reveal policy can accommodate nontrivial priorities, this subsection adds neighborhood priority to the first model.

	Formally, on the basis of the first model, we assume that, for each school $ s $, a subset $ I_s $ of students live in its neighborhood, with $ |I_s|=n_s\in (0,q_s) $. So, each school has an additional capacity to admit students outside of its neighborhood. Each student lives in the neighborhood of at most one school, and a number $ n-\sum_{s\in S}n_s $ of students do not live in any neighborhood. Each school $s$ uses a priority order in which the students in $ I_s $ are ranked above the others, and the students within $ I_s $ or $ I\backslash I_s $ are in ties. Before running $ \text{DA}^\ell $, priority ties are broken according to a randomly drawn lottery $ \pi: I\rightarrow \{1,2,\ldots,n\} $.

	Different from the common priority model, in the presence of neighborhood priority, informing each student of his lottery number is insufficient to resolve all uncertainties for students. Consider a student $ i $ whose lottery number is $ \pi(i)  $. In the common priority model, as long as $ \pi(i)\le  q_{s_1} $, $ i $ is sure of admission to $ s_1 $ by ranking $ s_1 $ highest in his ROL. However, in the present model, $ i $ is not sure of admission to $ s_1 $ by doing so, because all students in the neighborhood of $ s_1 $ must apply to $ s_1 $ and be admitted, irrespective of their lottery numbers. In other words, school $ s $ has only $ q_{s_1}-n_{s_1} $ seats for students outside its neighborhood. Only if $ \pi(i)\le q_{s_1}-n_{s_1} $, can $ i $ be sure of admission to $ s_1 $ by ranking it highest. If $  q_{s_1}-n_{s_1}<\pi(i) \le q_{s_1} $, $ i $ is uncertain about the number of students who receive worse lottery numbers than his but enjoy neighborhood priority for $ s_1$. To evaluate the admission probability, $ i $ needs to calculate the probability of events involving combinatorics.

	To deal with nontrivial priorities,  policymakers can provide a variety of statistics about the lottery outcome to reduce uncertainties for students. To illustrate this, for the neighborhood priority model in this subsection, we show that a more informative reveal policy than the one in the previous model can eliminate uncertainties for students and simplify their strategies.

	Specifically, we assume that, for any drawn lottery $ \pi $, policymakers privately reveal $ \pi(i) $ to each student $ i $ and publicly reveal the following statistics for the neighborhood of each school $s$: for each possible lottery number $ x$, policymakers publicly disclose the number of students from $I_s$ who receive lottery numbers weakly better than $ x $, that is, $ n^\pi_s(x)=|\{i\in I_s: \pi(i)\le x \}| $. 
	From these statistics, every student can infer the distribution of lottery numbers received by the students from each $ I_s $. We call this more informative disclosure policy the \textbf{RevealMore policy}.

	Under the RevealMore policy, for any realized utilities and drawn lottery, there exists a unique BNE, which is characterized by a vector of school cutoffs. Each student determines his optimal strategy based on his lottery number and the cutoffs, and is admitted to his best achievable school.
	
	\begin{proposition}\label{prop:neighborhood:reveal}
		In the neighborhood priority model, under the RevealMore policy, for any realized utilities and drawn lottery $ \pi $, there exists a unique BNE, characterized by a vector of school cutoffs $(x^\pi_1,x^\pi_2,\ldots,x^\pi_m)\in \{1,2,\ldots,n\}^m$,  with $x^\pi_1>x^\pi_2>\cdots>x^\pi_m $, such that,  for every student $i$:
		\begin{itemize}
			\item If $\pi(i)\in [1,x^\pi_1]$, or $i\in I_{s_1}$, then $i$ is admitted to $ s_1 $ by ranking it highest in his ROL;
			
			\item If, for some $2\le k\le m$, $\pi(i)\in (x^\pi_{k-1},x^\pi_k]$, or $\pi(i)>x^\pi_k$ but $i\in I_{s_k}$, then $i$ is admitted to $ s_k $ by ranking it  essentially highest in his ROL.
		\end{itemize}	
	\end{proposition}
	
	\begin{proof}
		In any BNE, every $ i\in I_{s_1} $ must be admitted to $ s_1 $ by ranking it highest. For any student $i\notin I_{s_1}$ who receives a lottery number $ x $, from the revealed lottery information, he knows that the number of students who receive lottery numbers worse than his yet have neighborhood priority for $s_1$ is $ n_{s_1}- n^\pi_{s_1}(x)$. Thus, as long as $x+n_{s_1}- n^\pi_{s_1}(x)\le q_{s_1}$, $i$ is sure of admission to $s_1$ by ranking it highest. So, the cutoff $ x^\pi_1 $ for school $ s_1  $ is the solution to 
		$$
		x+n_{s_1}- n^\pi_{s_1}(x)= q_{s_1}.$$

		Therefore, among the students not in the neighborhood of $ s_1 $, those whose lottery numbers are weakly better than $ x^\pi_1 $ must be admitted to $ s_1 $ by ranking it highest, and other students do not target $ s_1 $ because they know they cannot be admitted to $ s_1 $.
		
		Fixing the strategies of those students who are admitted to $s_1$, among the remaining students, every  $i\in I_{s_2}$ must be admitted to $s_2$ by ranking it highest (or only below $s_1$). For any lottery number $ x>x^\pi_1 $, the number of students who receive lottery numbers worse than $ x $ but have neighborhood priority for $s_2$ is $ n_{s_2}- n^\pi_{s_2}(x) $, while the number of students who receive lottery numbers in $(x^\pi_1,x] $ but have neighborhood priority for $ s_1 $ is $ n^\pi_{s_1}(x)- n^\pi_{s_1}(x^\pi_1) $. Thus, for any student $ i\notin I_{s_1}\cup I_{s_2} $ with a lottery number $ x > x^\pi_1$,  as long as $ [x-x^\pi_1]-[n^\pi_{s_1}(x)- n^\pi_{s_1}(x^\pi_1)]+[n_{s_2}- n^\pi_{s_2}(x)]\le q_{s_2} $, $ i $ is sure of admission to $ s_2 $ by ranking it highest (or only below $ s_1 $). So, the cutoff $ x^\pi_2 $ for school $ s_2 $ is the solution to
		$$
		[x-x^\pi_1]-[n^\pi_{s_1}(x)- n^\pi_{s_1}(x^\pi_1)]+[n_{s_2}- n^\pi_{s_2}(x)]= q_{s_2}.$$
		
		Therefore, among the students not in the neighborhoods of $ s_1 $ and $ s_2 $, those whose lottery numbers are in $(x^\pi_1,x] $ must be admitted to $s_2$ by ranking it essentially highest, and other students do not target $ s_2 $ because they know that they cannot be admitted to $ s_2 $.

		In general, for every school $ s_k $, with $ 1< k<m $, 	
		the cutoff $x^\pi_k$ is the solution to 
		$$(x-x^\pi_{k-1})- \sum_{y=1}^{k-1} [n^\pi_{s_y}(x)-n^\pi_{s_y}(x^\pi_{k-1})]+[n_{s_k}- n^\pi_{s_k}(x)]=q_{s_k},$$
		where $ \sum_{y=1}^{k-1} [n^\pi_{s_y}(x)-n^\pi_{s_y}(x^\pi_{k-1})] $ is the number of students who are in the neighborhood of better schools and receive lottery numbers in $(x^\pi_{k-1},x] $, and $ n_{s_k}- n^\pi_{s_k}(x) $ is the number of students who receive lottery numbers worse than $ x $ but have neighborhood priority for $s_k$.

		Because $s_m$ is the worst school and the total capacity of schools is exactly enough to admit all students, the cutoff for $s_m$ must be $x^\pi_m=n$, meaning that any student who cannot be admitted to better schools can be admitted to $ s_m $ by ranking it in their ROL.
		
		The above analysis implies that the equilibrium is unique. 
	\end{proof}

	Under the RevealMore policy, students can straightforwardly determine their best achievable school by observing their lottery numbers and the publicly disclosed statistics about the lottery. In the outcome, the number of students who attend schools outside their neighborhoods is weakly greater than $\sum_{k=1}^m (q_{s_k}-n_{s_k})$, and it equals $\sum_{k=1}^m (q_{s_k}-n_{s_k})$ only if all neighborhood students are unlucky in the drawn lottery such that they have to attend their neighborhood schools.\footnote{Specifically, the number of students admitted to $ s_1 $ but not in the neighborhood of $ s_1 $ is $x^\pi_1-n^\pi_{s_1}(x^\pi_1)$, which equals $q_{s_1}-n_{s_1}$. For every other school $s_k$, the number of students admitted to $ s_k $ but not in the neighborhood of $ s_k $ is $ (x^\pi_{k}-x^\pi_{k-1})- \sum_{y=1}^{k-1} [n^\pi_{s_y}(x^\pi_{k})-n^\pi_{s_y}(x^\pi_{k-1})] $, which is weakly greater than $q_{s_k}-n_{s_k}$, since $n_{s_k}-n^\pi_{s_k}(x^\pi_{k-1})\le n_{s_k}$ of its neighborhood students are admitted to $s_k$.}

	Under the Cover policy, however, since lottery outcomes are not revealed to students, except for the neighborhood students of $s_1$ who must be admitted to $s_1$ by ranking it highest, the remaining students must choose their strategies under uncertainty. Although it is generally difficult to characterize their equilibrium strategies, it is conceivable that, for those who have neighborhood schools, if the ROL length is significantly restricted and their utilities for their neighborhood schools are not significantly lower than for other schools, the uncertainty arising from the undisclosed lottery can lead them to choose to stay in their neighborhoods, which counteracts the purpose of school choice. For instance, suppose that the ROL length is one and students have utilities such that, in the equilibrium, all students who do not live in any neighborhood report schools strictly better than $s_m$, and they are distributed such that, for each $s_k$ with $ k<m $, a number $z_{s_k}$ of them report $s_k$, with $z_{s_k}> q_{s_k}-n_{s_k}$.\footnote{This is possible because $\sum_{k=1}^{m-1} (q_{s_k}-n_{s_k})=n-q_{s_m}-\sum_{k=1}^{m-1}n_{s_k}<n-n_{s_m}-\sum_{k=1}^{m-1} n_{s_k}=\sum_{k=1}^{m-1}z_{s_k}$.}  Then, every student who has a neighborhood school must report his neighborhood school in his ROL and stay in the neighborhood if their utilities satisfy the following condition: for every two schools $s_y$ and $s_u$ with $u<y$ and for every $i\in I_{s_y}$, $v^i_y>\frac{q_{s_u}-n_{s_u}}{z_{s_u}} v^i_u$.\footnote{Since all neighborhood students of $s_1$ must report $s_1$ and be admitted, if any other student reports $s_1$, the probability of attending $s_1$ is no more than $\frac{q_{s_1}-n_{s_1}}{z_{s_1}}$. So, for any $i\in I_{s_y}$ with $y>1$, if $v^i_y>\frac{q_{s_1}-n_{s_1}}{z_{s_1}} v^i_1$, then reporting $s_1$ is worse than reporting $s_y$. Thus, all neighborhood students of $s_2$ must report $s_2$. Given this, if any other student reports $s_2$, the probability of attending $s_2$ is no more than $\frac{q_{s_2}-n_{s_2}}{z_{s_2}}$. So, for any $i\in I_{s_y}$ with $y>2$, if $v^i_y>\frac{q_{s_2}-n_{s_2}}{z_{s_2}} v^i_2$, then reporting $s_2$ is worse than reporting $s_y$. Thus, all neighborhood students of $s_3$ must report $s_3$. In general, if,  for every $s_y$ and $s_u$ with $u<y$ and for every $i\in I_{s_y}$, $v^i_y>\frac{q_{s_u}-n_{s_u}}{z_{s_u}} v^i_u$, then $i$ must report his neighborhood school in the equilibrium.} In this equilibrium, the capacity of school $s_m$ is wasted, and an equal number of students are unmatched.

	We present a modification of Example \ref{example:first:model} to illustrate these results. The environment in the example is also used in our experiment.

	\begin{example}\label{example:second:model}
		On the basis of Example \ref{example:first:model}, we further assume that the three students $1,2,3$ respectively live in the neighborhoods of $s_1,s_2,s_3$. The others do not live in any neighborhood. We analyze the three policies regarding the lottery information.
		
		\textbf{Cover policy}: There is a unique symmetric BNE in which students $1$ and $2$ respectively report their neighborhood schools, $ s_1 $ and $ s_2 $. Among the remaining students, every type-$ \bv $ student reports $ s_1 $ and every type-$ \bv' $ student reports $ s_2 $. No students report $s_3$.\footnote{In Example \ref{example:first:model}, we do not set a different utility type for student $3$. As a result, in Example \ref{example:second:model}, student $3$ plays the same strategy as those without neighborhoods under the Cover policy. If we set a relatively high utility of school $s_3$ for student $3$, then student $3$ would report his neighborhood school in the equilibrium.} So, $ s_3 $ is always vacant, and at least two students are unmatched.
		
		From an ex-ante view, the expected match rate is about $63\%$. Students $1$ and $2$ must be admitted to their neighborhood schools. So, the expected utility is about $83.33$  for student $1$ and about $46.67$ for student $2$. For others, from an interim view, every type-$ \bv $ student is admitted to $ s_1 $ with probability $\frac{10}{27}\approx 0.37$, and every type-$ \bv' $ student is admitted to $ s_2 $ with probability $\frac{65}{108}\approx 0.60$. From an ex-ante view, they receive the (approximate) random assignment $(0.25s_1,0.20s_2,0.55\emptyset)$, and their expected utility is about $34.26$.
		
		\textbf{Reveal policy}: Students observe their own lottery numbers but not others'. In equilibrium, students' strategies depend on both their own lottery numbers and their neighborhood schools. Student $1$ always reports $ s_1 $. Student $ 2 $ reports $ s_1 $ if his lottery number is 1; otherwise, he reports $ s_2 $. Student $ 3 $ reports $s_1$ if his lottery number is 1; he reports $ s_2 $ if his lottery number is 2 or 3; in other cases, he reports $s_3$. Each no-neighborhood student’ strategy is similar to student $ 3 $ but with one exception: he reports $ s_1 $ if his lottery number is 5. That is, no-neighborhood students’ strategies are not ``monotone'' in their lottery numbers. When their lottery number goes from 1 to 4, they report a weakly worse school as their lottery number gets worse, until reporting $s_3$ when their lottery number is 4. When their lottery number is greater than 4, they reset the monotone strategy: they report $s_1$ when their lottery number is 5 and report $s_3$ when their lottery number is 6. Overall, students $1$ and $2$ must be admitted to their reported schools, while the other students cannot be assured admission.
		
		From an ex-ante view, the expected match rate is about $87\%$. Student $1$ must be admitted to $s_1$. His expected utility is $83.33$. The expected assignment is $(\frac{1}{6}s_1,\frac{5}{6}s_2)$ for student $2$, $(\frac{1}{6}s_1,\frac{31}{120}s_2,\frac{1}{2}s_3,\frac{3}{40}\emptyset)$ for student $3$, and $(\frac{1}{5}s_1,\frac{31}{120}s_2,\frac{19}{60}s_3,\frac{9}{40}\emptyset)$ for every other student. So, the expected utility is about $52.78$ for student $2$, $35.94$ for student $3$, and $35.06$ for every other student.
		
		\textbf{RevealMore policy}: 
		Students observe both their own lottery numbers and the lottery numbers of students in all neighborhoods. This is a special case of the RevealMore policy defined above, since there is only one student in each school's neighborhood. In the equilibrium, after observing the lottery information, every student can predict any other students' strategies and then determine their best achievable school. This ensures that every student is admitted to his reported school. 
		
		Specifically, student $1$ always reports $s_1$. Student $2$ reports $s_1$ if his lottery number is 1, or if his lottery number is 2 and he observes that student $1$'s lottery number is 1. Under these two cases, student $2$ can ensure admission to $s_1$. In all other cases, student $2$ reports $s_2$. All other students' strategies follow a similar logic: they report $s_1$ if their lottery number is 1, or if their lottery number is 2 and they observe that student $1$'s lottery number is 1. They report $s_2$ if a) their lottery number is 2 and student $1$'s lottery number is greater than 1, or if b) their lottery number is 3 and student $1$'s lottery number is lower than 3, or if c) their lottery number is 3 and student $1$'s lottery number is greater than 2 and student $2$'s lottery number is lower than 3, or if d) their lottery number is 4 and both student $1$'s and student $2$'s lottery numbers are lower than 4. In all other cases, they report $s_3$.
		
		
		From an ex-ante view, the expected match rate is $100\%$. Student $1$ must be admitted to $s_1$, so his expected utility is $83.33$. The expected assignment is $(\frac{1}{5}s_1,\frac{4}{5}s_2)$ for student $2$, and  $(\frac{1}{5}s_1,\frac{3}{10}s_2,\frac{1}{2}s_3)$ for each other student. So, the expected utility is $54$ for student $2$, and about $40.67$ for every other student. 
		
		Hence, across the three policies, as policymakers reveal more detailed lottery information, all students but student $1$ obtain strictly higher expected match rates and expected utilities.
		
	\end{example}
	
	Our model focuses specifically on neighborhood priority, but in practice, school choice systems often incorporate multiple priority classes. In such cases, implementing the RevealMore policy would require policymakers to reveal more detailed information about the lottery, which might be challenging for policymakers due to operational constraints and might overwhelm students with excessive complexity in digesting such information. To conclude this subsection, we note that in practical markets, there may exist other key forces that can significantly simplify the reveal policy under nontrivial priorities. 
	
	First, the law of large numbers in sufficiently large markets reduces uncertainty about lottery number distributions across student groups. While our finite market model shows that cutoffs vary with lottery draws (Proposition \ref{prop:neighborhood:reveal}), a continuum approximation demonstrates how cutoffs stabilize in large markets. Consider a continuum model where student mass is normalized to 1, and each school $ s $ has capacity $ q_s\in (0,1) $ and neighborhood mass $ n_s\in (0,q_s) $. A lottery draw is represented by a one-to-one mapping $ \pi: I \rightarrow [0,1] $. Then, for any school $ s $, the mass of its neighborhood students who receive lottery numbers within any interval $ [x_1,x_2] $ becomes deterministic: $ (x_2-x_1)n_{s} $. This leads to stable equilibrium cutoffs $ x^\pi_k=\frac{\sum_{y=1}^kq_{s_y}-\sum_{y=1}^kn_{s_y}}{1-\sum_{y=1}^kn_{s_y}}$, where lotteries only determine which specific students are admitted without affecting the cutoff values. Therefore, for this continuum model, revealing to each student his lottery number is sufficient to choose the optimal strategy.

	Second, historical information provides another simplifying force when market fundamentals remain stable across periods. In such cases, policymakers may reveal historical cutoffs that can serve as reliable reference points for students to estimate their chances and choose their optimal strategies.
	
	These observations suggest that practical implementations may not require the full information revelation implied by our model. Policymakers can implement simpler revelation policies based on observed market stability across periods and market size within each period, while still providing students with sufficient information when necessary to simplify their strategies.

	\section{Experiments}\label{section:experiment}
	
	To evaluate the advantages of the Reveal and RevealMore policies compared to the Cover policy, we conduct laboratory experiments involving school choice markets both with and without neighborhood schools. Sections \ref{sec:design} and \ref{sec:results} present the design, hypotheses and results of the main experiment, which closely follows the setup of Examples \ref{example:first:model} and \ref{example:second:model}, thereby enabling rigorous testing of the theoretical predictions. Section \ref{sec:exp2} briefly discusses a robustness experiment that shares the primary research objectives but introduces a more complex experimental environment by allowing students to rank multiple schools in their preference lists.

	\subsection{Experimental Design}\label{sec:design}
	
	\paragraph{Environment} 
	We design a school choice market involving 6 students (ID1 to ID6) and 3 schools ($s_1$, $s_2$, $s_3$). Each school can admit 2 students. Students are played by experimental participants, while schools are simulated by the computer.
	Students have identical ordinal preferences $ s_1 \succ s_2 \succ s_3 $, but they may differ in their assigned cardinal utilities (i.e., payoffs). Specifically, there are two types of utilities $ \{\bv,\bv'\} $. Each student's type is independently drawn from the distribution $ f(\bv)=2/3 $ and $ f(\bv')=1/3 $. Each student can report one school in his ROL.
	\begin{table}[!htb]
		\centering
		\begin{tabular}{|c|c|c|}
			\hline
			& $ \bv $ & $ \bv' $ \\ \hline
			$ s_1 $ & $ 90 $ &  $ 70 $ \\ \hline
			$ s_2 $ & $ 40 $ &  $ 60 $ \\ \hline
			$ s_3 $ & $ 20 $ &  $ 20 $ \\ \hline			
		\end{tabular}
	\end{table}

	\paragraph{Treatments} 
	We implement a total of five treatments using a between-subjects design. The first four treatments follow a 2$\times$2 factorial design, varying both the availability of neighborhood schools and the lottery revelation policy. In the two no-neighborhood treatments, referred to as NoNS\_Cover and NoNS\_Reveal, all students are in priority ties for all schools. A single lottery is drawn uniformly at random to resolve these priority ties. Subsequently, the standard deferred acceptance mechanism is employed to determine the matching outcome. In the two neighborhood treatments, termed NS\_Cover and NS\_Reveal, each school has one student living in its neighborhood. Specifically, student ID1 lives in the neighborhood of school $s_1$ and is thus designated as the $s_1$-neighbor. Similarly, students ID2 and ID3 live in the neighborhood of schools $s_2$ and $s_3$, respectively, and are referred to as the $s_2$-neighbor and $s_3$-neighbor. These students hold the unique highest priority for their respective neighborhood schools. All other students are in priority ties, which are resolved by a single random lottery. 
	
	The second experimental dimension manipulated is the lottery revelation policy. In the NoNS\_Cover and NS\_Cover treatments, the Cover policy is implemented, whereby students do not learn their lottery numbers before submitting their preferences. In contrast, the NoNS\_Reveal and NS\_Reveal treatments implement the Reveal policy, in which students observe their own lottery number before submitting preferences. 
	
	The fifth treatment, called NS\_RevealMore is motivated by the theoretical result presented in Subsection \ref{theory:neighbor} and implements the RevealMore policy within the same neighborhood school setup. Compared to the Reveal policy, each student observes not only their own lottery number but also the lottery numbers of all neighborhood students. \autoref{tab:design} summarizes the experimental design alongside aggregate-level theoretical predictions.

	\begin{table}[!htb]
		\centering
		\caption{Experimental design and aggregate-level theoretical predictions}
		\label{tab:design}
		\begin{adjustbox}{width=1\textwidth}
			\begin{tabular}{lccccc}
				\hline
				Treatment	            & NoNS\_Cover& NoNS\_Reveal& NS\_Cover  & NS\_Reveal  & NS\_RevealMore \\ 
				\hline
				Policy                & Cover      & Reveal      & Cover      & Reveal      & RevealMore     \\
				Neighborhood school   & No         & No          & Yes        & Yes         & Yes           \\
				&            &             &            &             &               \\
				Expected match rate   & 59\%       & 100\%       & 63\%       & 87\%        & 100\%         \\ 
				Expected match rate for type-$\bv$ & 49\% & 100\% &  & & \\
				Expected match rate for type-$\bv'$& 78\% & 100\% &  & & \\
				Expected match rate for $s_1$-neighbor &      &  & 100\% & 100\% & 100\% \\
				Expected match rate for $s_2$-neighbor &      &  & 100\% & 100\% & 100\% \\
				Expected match rate for $s_3$-neighbor &      &  & 44.67\% & 92.50\%& 100\% \\
				Expected match rate for others         &      &  & 44.67\% & 77.50\%& 100\% \\
				&            &             &            &             &               \\
				Expected payoff       & 45.33      & 50          & 44.50      & 46.21       & 50            \\
				Expected payoff for type-$\bv$         & 44.57 & 50 & & & \\
				Expected payoff for type-$\bv'$        & 46.86 & 50 & & & \\
				Expected payoff for $s_1$-neighbor     & &       & 83.33      & 83.33       & 83.33         \\ 
				Expected payoff for $s_2$-neighbor     & &       & 46.67      & 52.78       & 54            \\ 
				Expected payoff for $s_3$-neighbor     & &       & 34.26      & 35.94       & 40.67         \\ 
				Expected payoff for others             & &       & 34.26      & 35.06       & 40.67         \\
				&            &             &            &             &               \\
				Sessions              & 4          & 4           & 4          & 4           & 4             \\
				Participants          & 48         & 48          & 48         & 48          & 48            \\
				\hline			
			\end{tabular}
		\end{adjustbox}
	\end{table}

	\paragraph{Theoretical Predictions}
	
	Equilibrium strategies for each treatment have been thoroughly analyzed in Examples \ref{example:first:model} and \ref{example:second:model}, and are summarized in \autoref{tab:equilibrium-prediction}. Based on these equilibrium strategies, we derive theoretical predictions for the (interim) expected match rates and (interim) expected payoffs associated with each treatment, as presented in \autoref{tab:design}. Additionally, we provide predictions regarding the frequencies of submission strategies contingent upon students' roles and types within each treatment, which are detailed in \autoref{tab:strategy-prediction}.

	\begin{table}[!htb]
		\centering
		\caption{Equilibrium strategies in each treatment} 
		\label{tab:equilibrium-prediction}    
		\begin{subtable}[t]{\textwidth}
			\centering
			\caption{NoNS\_Cover and NS\_Cover}
			\begin{tabular}{cccccccc}
				\hline
				\multicolumn{2}{c}{NoNS\_Cover}  & \multicolumn{6}{c}{NS\_Cover} \\ \cmidrule(lr){1-2} \cmidrule(lr){3-8} 
				Type-$\bv$ & Type-$\bv'$         & $ s_1 $-neighbor  & $ s_2 $-neighbor & \multicolumn{2}{c}{$ s_3 $-neighbor} & \multicolumn{2}{c}{others} \\ \cmidrule(lr){5-6} \cmidrule(lr){7-8}
				&                     &                   &                  &  Type-$\bv$ & Type-$\bv'$ &  Type-$\bv$ & Type-$\bv'$ \\ \hline
				$ s_1 $    & $ s_2 $             & $ s_1 $           & $ s_2 $          & $ s_1 $ & $ s_2 $ & $ s_1 $ & $ s_2 $ \\ 
				\hline 
			\end{tabular}
		\end{subtable}
		\hfill
		
		\begin{subtable}[t]{\textwidth}
			\centering
			\caption{NoNS\_Reveal and NS\_Reveal}
			\begin{tabular}{lccccc}
				\hline
				\multirow{2}{*}{lottery info} & NoNS\_Reveal & \multicolumn{4}{c}{NS\_Reveal} \\ \cmidrule(lr){2-2} \cmidrule(lr){3-6} 
				& all students & $ s_1 $-neighbor & $ s_2 $-neighbor & $ s_3 $-neighbor & others \\ \hline
				own lottery = 1 & $ s_1 $ & $ s_1 $ & $ s_1 $ & $ s_1 $ & $ s_1 $ \\ 
				own lottery = 2 & $ s_1 $ & $ s_1 $ & $ s_2 $ & $ s_2 $ & $ s_2 $ \\ 
				own lottery = 3 & $ s_2 $ & $ s_1 $ & $ s_2 $ & $ s_2 $ & $ s_2 $ \\ 
				own lottery = 4 & $ s_2 $ & $ s_1 $ & $ s_2 $ & $ s_3 $ & $ s_3 $ \\ 
				own lottery = 5 & $ s_3 $ & $ s_1 $ & $ s_2 $ & $ s_3 $ & $ s_1 $ \\ 
				own lottery = 6 & $ s_3 $ & $ s_1 $ & $ s_2 $ & $ s_3 $ & $ s_3 $ \\ \hline
			\end{tabular}
		\end{subtable}
		\hfill
		
		\begin{subtable}[t]{\textwidth}
			\centering
			\caption{NS\_RevealMore}
			\begin{tabular}{lcccc}
				\hline 
				lottery info & $ s_1 $-neighbor & $ s_2 $-neighbor & $ s_3 $-neighbor & others \\ \hline
				own lottery = 1 & $ s_1 $ & $ s_1 $ & $ s_1 $ & $ s_1 $ \\ 
				own lottery = 2 \& & \multirow{2}{*}{$ s_1 $} & \multirow{2}{*}{$ s_1 $} & \multirow{2}{*}{$ s_1 $} & \multirow{2}{*}{$ s_1 $} \\
				\hspace{10pt} $ s_1 $-neighbor's lottery = 1 \\ 
				own lottery = 2 \& & \multirow{2}{*}{$ s_1 $} & \multirow{2}{*}{$ s_2 $} & \multirow{2}{*}{$ s_2 $} & \multirow{2}{*}{$ s_2 $} \\
				\hspace{10pt} $ s_1 $-neighbor's lottery > 1 \\ 
				own lottery = 3 \& & \multirow{2}{*}{$ s_1 $} & \multirow{2}{*}{$ s_2 $} & \multirow{2}{*}{$ s_2 $} & \multirow{2}{*}{$ s_2 $} \\
				\hspace{10pt} $ s_1 $-neighbor's lottery $\le$ 2 \\ 
				own lottery = 3 \& & \multirow{3}{*}{$ s_1 $} & \multirow{3}{*}{$ s_2 $} & \multirow{3}{*}{$ s_2 $} & \multirow{3}{*}{$ s_2 $} \\ 
				\hspace{10pt} $ s_1 $-neighbor's lottery > 2 \& &  \\
				\hspace{10pt} $ s_2 $-neighbor's lottery $\le 2$ \\ 
				own lottery = 4 \& & \multirow{3}{*}{$ s_1 $} & \multirow{3}{*}{$ s_2 $} & \multirow{3}{*}{$ s_2 $} & \multirow{3}{*}{$ s_2 $} \\ 
				\hspace{10pt} $ s_1 $-neighbor's lottery $\le 3$ \& &  \\
				\hspace{10pt} $ s_2 $-neighbor's lottery $\le 3$ \\ 
				other cases & $ s_1 $ & $ s_2 $ & $ s_3 $ & $ s_3 $ \\ \hline
			\end{tabular}
		\end{subtable}
	\end{table}

	\begin{table}[!htb]
		\caption{Prediction of frequencies of submission strategies (\%)}
		\label{tab:strategy-prediction}
		\begin{center}
			\begin{adjustbox}{width=1\textwidth}
				\begin{tabular}{lcccccccccc}
					\hline
					& \multicolumn{5}{c}{Type-$\bv$}                                      & \multicolumn{5}{c}{Type-$\bv'$}                                 \\
					\cmidrule(lr){2-6} \cmidrule(lr){7-11}
					Strategy	      & all     & $s_1$-neighbor & $s_2$-neighbor & $s_3$-neighbor & other  & all     & $s_1$-neighbor & $s_2$-neighbor & $s_3$-neighbor & other  \\
					\hline
					\multicolumn{11}{l}{\textbf{NoNS\_Cover}} \\
					$s_1$           & 100   & & & &                                                   & 0   & & & &  \\
					$s_2$           & 0   & & & &                                                     & 100   & & & &  \\
					$s_3$           & 0    & & & &                                                    & 0    & & & &  \\
					\multicolumn{11}{l}{\textbf{NoNS\_Reveal}} \\
					$s_1$           & 33.33   & & & &                                                 & 33.33    & & & &  \\
					$s_2$           & 33.33    & & & &                                                & 33.33    & & & &  \\
					$s_3$           & 33.33    & & & &                                                & 33.33    & & & &  \\
					\multicolumn{11}{l}{\textbf{NS\_Cover}} \\
					$s_1$           &    & 100            & 0              & 100            & 100     &    & 100      & 0            & 0           & 0 \\
					$s_2$           &    & 0              & 100            & 0              & 0       &    & 0        & 100          & 100         & 100 \\
					$s_3$           &    & 0              & 0              & 0              & 0       &    & 0        & 0            & 0           & 0 \\
					\multicolumn{11}{l}{\textbf{NS\_Reveal}} \\
					$s_1$           &    & 100            & 16.67          & 16.67          & 33.33   &    & 100      & 16.67        & 16.67       & 33.33 \\
					$s_2$           &    & 0              & 83.33          & 33.33          & 33.33   &    & 0        & 83.33        & 33.33       & 33.33 \\
					$s_3$           &    & 0              & 0              & 50.00          & 33.33   &    & 0        & 0            & 50.00       & 33.33 \\
					\multicolumn{11}{l}{\textbf{NS\_RevealMore}} \\
					$s_1$           &    & 100            & 20             & 20             & 20      &    & 100      & 20           & 20          & 20 \\
					$s_2$           &    & 0              & 80             & 30             & 30      &    & 0        & 80           & 30          & 30 \\
					$s_3$           &    & 0              & 0              & 50             & 50      &    & 0        & 0            & 50          & 50 \\
					\hline			
				\end{tabular}
			\end{adjustbox}
		\end{center}
		\par
		\vspace{-0.3cm}{\footnotesize \textit{Notes:}  }  
		\par
	\end{table}

	\paragraph{Hypotheses}
	Drawing upon the theoretical predictions, we formulate and test the following key hypotheses:
	
	\begin{hypothesis}\label{exp1:hypothesis:1} 
		The overall match rate and average student payoff are ordered across treatments as follows: NoNS\_Reveal > NoNS\_Cover and NS\_RevealMore > NS\_Reveal > NS\_Cover.
	\end{hypothesis}
	
	\begin{hypothesis}\label{exp1:hypothesis:2} 
		For the two no-neighborhood treatments, the Reveal policy improves the welfare of students of both utility types. For the three neighborhood treatments, the Reveal and RevealMore policies improve the welfare of all students but $s_1$-neighbor. 
	\end{hypothesis}
	
	\begin{hypothesis}\label{exp1:hypothesis:3} 
		At the individual level, participants' likelihood to play equilibrium strategies is ordered as follows: NoNS\_Reveal > NoNS\_Cover and NS\_RevealMore > NS\_Reveal > NS\_Cover.
	\end{hypothesis}

	\paragraph{Experimental Procedure}
	The experiment was conducted in May 2025 at the Economics Experimental Lab of Nanjing Audit University with a total of 240 university students, using the software z-Tree \citep{Fischbacher2007}. We ran four sessions for each treatment. Each session consisted of 12 participants who interacted for a total of 20 rounds. In each round, participants were randomly assigned to two 6-person groups. Within each group, each participant was randomly assigned a student ID (ranging from 1 to 6), a lottery number (also ranging from 1 to 6), and payoffs associated with attending each of the three schools; all of these values were distinct in each round. Specifically, we generated a sequence of the 5-tuples, consisting of the student ID, lottery number, and three payoffs, at the group-round level. We applied the sequence to each session to alleviate concerns that treatment effects might arise solely from differences in random number generation. After each round, participants received feedback about whether and to which school they were admitted and their round payoff. At the end of the session, one round was privately and randomly chosen for each participant to determine their final payoff. The experimental instructions are available in Online Appendix \ref{appendix:instructions}.
	
	Upon arrival, participants were randomly seated at partitioned computer terminals. The experimental instructions were given to participants in printed form and read aloud by the experimenter. The instructions contained a detailed example illustrating how the matching works. Participants then completed a comprehension quiz before proceeding to the experiment. At the end of the session, they filled out a demographic questionnaire. A typical session lasted about one hour with average earnings of 61.1 RMB, including a show-up fee of 15 RMB.\footnote{The average hourly earnings from the experiment were considerably higher than the local minimum wage of about 15-20 RMB per hour.}

	\subsection{Experimental Results}\label{sec:results}
	
	\paragraph{Match Rate} 
	\autoref{tab:summary} reports the observed match rate for each treatment. The Reveal policy yields a significantly higher match rate than the Cover policy, irrespective of the presence of neighborhood schools. In the absence of neighborhood schools, the average match rate under the Reveal policy is 96.25\%, which is significantly higher than 67.81\% under the Cover policy ($p$ = 0.029, Wilcoxon rank-sum test with each session treated as an independent observation; same below).\footnote{With 4 sessions per treatment, 0.029 is the lowest possible p-value when conducting this test. } Similarly, when neighborhood schools are available, the match rates are 93.54\% and 87.08\% under the two reveal policies (NS\_RevealMore and NS\_Reveal, respectively), compared to 77.92\% under the Cover policy ($p$ = 0.029). Importantly, the additional information about the lottery numbers of students with neighborhood schools provided in NS\_RevealMore significantly improves the match rate relative to the standard Reveal policy in NS\_Reveal ($p$ = 0.029).\footnote{Appendix \autoref{fig:match-rate} illustrates the match rate across rounds for each treatment, suggesting that the observed differences persist throughout the experiment.} These findings are consistent with Hypothesis \ref{exp1:hypothesis:1}. 
	
	Further, \autoref{tab:summary} indicates that, in the absence of neighborhood schools, the Reveal policy significantly improves the match rate for students of both utility types (type-$\bv$: 96.43\% vs. 67.70\%; type-$\bv'$: 95.89\% vs. 68.04\%; $p$ = 0.029 in both comparisons).  When neighborhood schools are present, while $s_1$-neighbor and $s_2$-neighbor students are almost always matched regardless of the lottery revelation policy, the Reveal policy significantly increases the match rate for $s_3$-neighbor students (92.50\% vs. 80.00\%, $p$ = 0.029) and for no-neighborhood students (78.13\% vs. 64.58\%, $p$ = 0.029). Additionally, the RevealMore policy further enhances the match rate for no-neighborhood students compared to the Reveal policy (89.79\% vs. 78.13\%, $p$ = 0.029). Although the RevealMore policy does not yield a statistically significant improvement for $s_3$-neighbor students, their match rate under the standard Reveal policy is already very high, limiting the potential for further significant gains. Collectively, these findings support Hypothesis \ref{exp1:hypothesis:2}.

	\begin{table}[!htb]
		\caption{Match rate and student payoff}
		\label{tab:summary}
		\begin{center}
			\begin{adjustbox}{width=1\textwidth}
				\begin{tabular}{lccccc}
					\hline
					Treatment	                         & NoNS\_Cover& NoNS\_Reveal& NS\_Cover  & NS\_Reveal  & NS\_RevealMore \\ 
					\hline
					Match rate                         & 67.81\%     & 96.25\%      & 77.92\%     & 87.08\%      & 93.54\%         \\ 
					& (1.51)      & (0.61)       & (1.34)      & (1.08)       & (0.79)          \\
					Match rate for type-$\bv$          & 67.70\%     & 96.43\%      &             &              &                 \\
					& (1.84)      & (0.73)       &             &              &                 \\
					Match rate for type-$\bv'$         & 68.04\%     & 95.89\%      &             &              &                 \\
					& (2.63)      & (1.12)       &             &              &                 \\
					Match rate for $s_1$-neighbor      &             &              & 100\%       & 100\%        & 100\%           \\
					&             &              & (0.00)      & (0.00)       & (0.00)          \\
					Match rate for $s_2$-neighbor      &             &              & 93.75\%     & 95.63\%      & 96.88\%         \\
					&             &              & (1.92)      & (1.62)       & (1.38)          \\
					Match rate for $s_3$-neighbor      &             &              & 80.00\%     & 92.50\%      & 95.00\%         \\
					&             &              & (3.17)      & (2.09)       & (1.73)          \\
					Match rate for others              &             &              & 64.58\%     & 78.13\%      & 89.79\%         \\
					&             &              & (2.19)      & (1.89)       & (1.38)          \\
					Average payoff                     & 42.27       & 48.72        & 43.53       & 44.78        & 47.81           \\
					& (1.15)      & (0.92)       & (1.06)      & (0.99)       & (0.94)          \\
					Average payoff for type-$\bv$      & 43.43       & 48.85        &             &              &                 \\
					& (1.50)      & (1.21)       &             &              &                 \\
					Average payoff for type-$\bv'$     & 39.91       & 48.45        &             &              &                 \\
					& (1.68)      & (1.29)       &             &              &                 \\
					Average payoff for $s_1$-neighbor  &             &              & 82.75       & 82.69        & 82.75           \\ 
					&             &              & (0.76)      & (0.77)       & (0.76)          \\ 
					Average payoff for $s_2$-neighbor  &             &              & 44.25       & 45.69        & 47.25           \\ 
					&             &              & (1.27)      & (1.22)       & (1.24)          \\ 
					Average payoff for $s_3$-neighbor  &             &              & 29.63       & 35.81        & 38.44           \\ 
					&             &              & (2.26)      & (2.15)       & (2.14)          \\ 
					Average payoff for others          &             &              & 34.85       & 34.83        & 39.48           \\
					&             &              & (1.53)      & (1.38)       & (1.32)          \\ 
					\hline			
				\end{tabular}
			\end{adjustbox}
		\end{center}
		\par
		\vspace{-0.3cm}{\footnotesize \textit{Notes:} Standard errors are in parentheses.}  
		\par
	\end{table}

	\paragraph{Efficiency} 
	\autoref{tab:summary} also reports the efficiency, measured by the average student payoff, for each treatment. In the absence of neighborhood schools, the Reveal policy yields significantly greater overall efficiency than the Cover policy (48.72 vs. 42.27, $p$ = 0.029); and this improvement holds for students of both utility types ($p$ = 0.029 in both cases). In the presence of neighborhood schools, the Reveal policy does not  produce a statistically significant increase in overall efficiency compared to the Cover policy (44.78 vs. 43.53, $p$ = 0.486). However, the RevealMore policy achieves a larger and statistically significant improvement over the Cover policy (47.81 vs. 43.53, $p$ = 0.029). Further, while the Reveal policy significantly increases overall efficiency for $s_3$-neighbor students (35.81 vs. 29.63, $p$ = 0.057) but not for no-neighborhood students (34.83 vs. 34.85, $p$ = 0.886) compared to the Cover policy, the RevealMore policy significantly enhances efficiency for both $s_3$-neighbor (38.44 vs. 35.81, $p$ = 0.029) and no-neighborhood students (39.48 vs. 34.83, $p$ = 0.057) compared to the Reveal policy. Although neither Reveal nor RevealMore policies significantly increase efficiency for $s_2$-neighbor students, the observed effects trend in the expected direction. Overall, these findings regarding efficiency parallel the results on match rates and are largely consistent with Hypotheses \ref{exp1:hypothesis:1} and \ref{exp1:hypothesis:2}.

\begin{figure}[!htp]
		\caption{Average absolute deviation }
		\begin{center}
			\begin{subfigure}[b]{0.47\textwidth}
				\centering
				\caption{Average}
				\label{fig:deviation}
				\includegraphics[width=\linewidth]{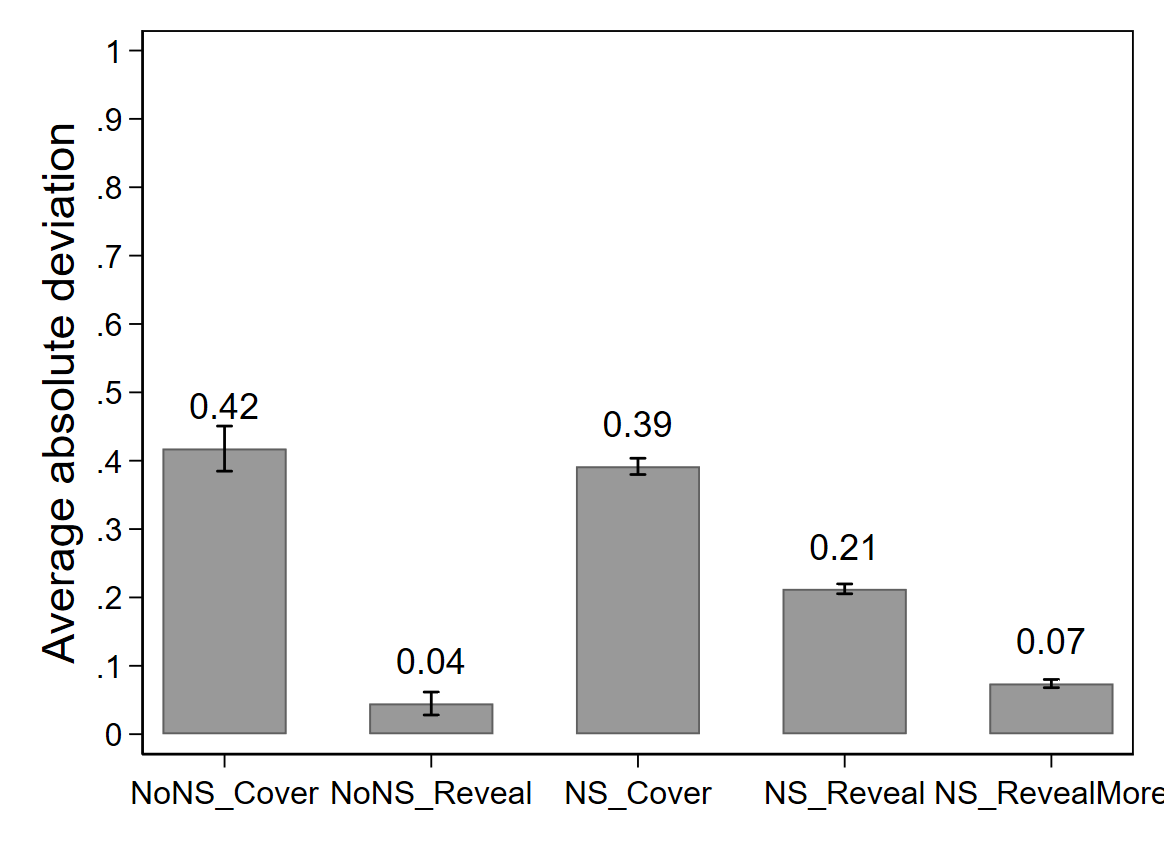}
			\end{subfigure}
			\begin{subfigure}[b]{0.47\textwidth}
				\centering
				\caption{Across rounds}
				\label{fig:deviation-round}
				\includegraphics[width=\linewidth]{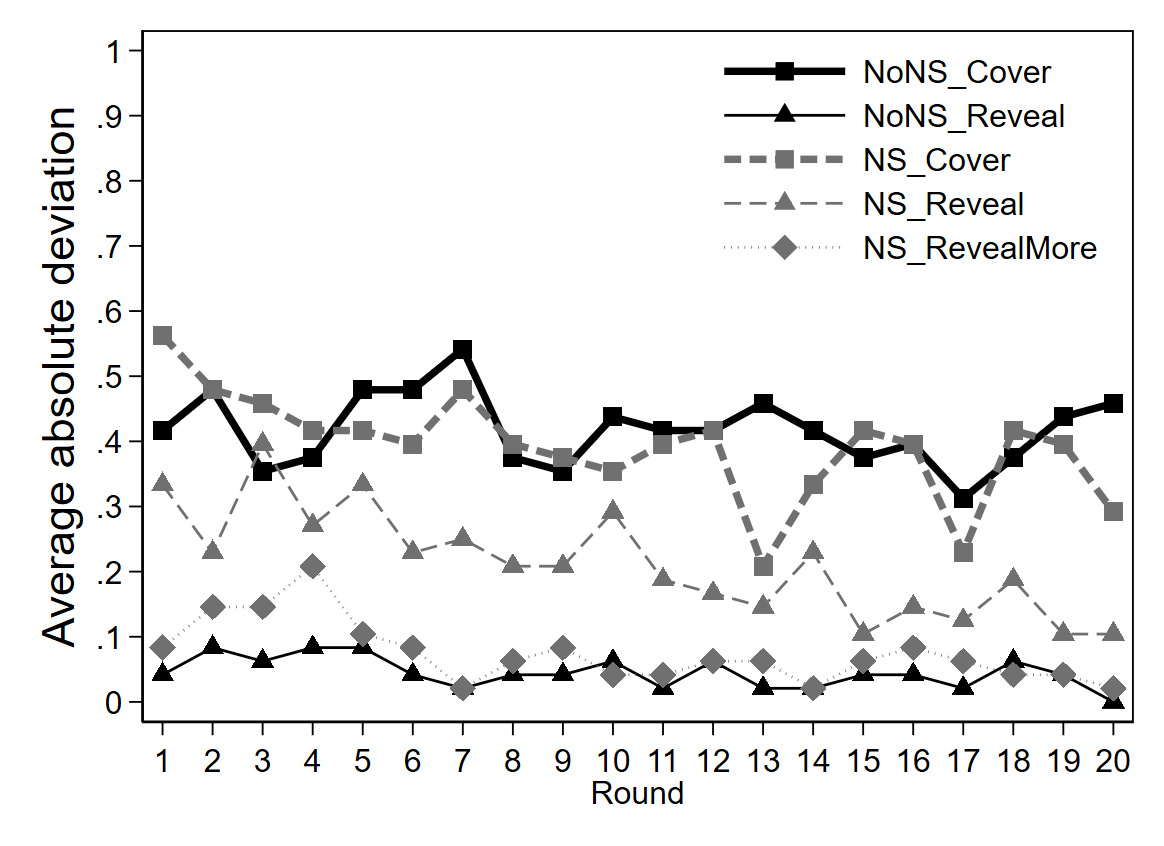}
			\end{subfigure}
		\end{center}
		\label{fig:aad}
		\par
		\vspace{-0.5cm}  {\footnotesize \textit{Notes:} Standard errors clustered at the matching group level are indicated by bars. }
	\end{figure}

	\paragraph{Individual Strategy} 
	To summarize participants’ adherence to equilibrium strategies across treatments, we construct a statistic termed average absolute deviation (AAD), which calculates the treatment-level frequency of deviating from the equilibrium strategies summarized in \autoref{tab:equilibrium-prediction}. \autoref{fig:aad} displays the results, showing that the Reveal policy substantially reduces AAD irrespective of the presence of neighborhood schools. In the absence of neighborhood schools, the Reveal policy reduces AAD from 0.42 to 0.04 ($p$ = 0.029), meaning that only 4\% of decisions deviate from equilibrium strategies. When neighborhood schools are available, the Reveal policy reduces AAD from 0.39 to 0.21, and the RevealMore policy further reduces it to 0.07 ($p$ = 0.029 in all treatment comparisons). \autoref{fig:aad} further shows that treatment differences persist across rounds, although AAD generally decreases over time, reflecting learning effects; this reduction is especially pronounced in NS\_Reveal.\footnote{Appendix \autoref{tab:aad-regression} reports complementary regression analyses confirming that treatment differences are significant both across all rounds and during the last 5 rounds, with a noticeable decline in suboptimal decisions in NS\_Reveal during the last 5 rounds. }

	Note that in NS\_Reveal, no-neighborhood students' strategies are ``non-monotonic'' in their lottery numbers. Specifically, their equilibrium strategy is to report $s_1$ when the lottery number is 5, which is unintuitive. Indeed, experimental data indicate that participants deviated from this strategy over 95\% of the time under this circumstance (see \autoref{tab:equilibrium}). Therefore, we performed a robustness check by excluding observations for no-neighborhood students with lottery number 5 across all treatments (to ensure fair comparisons). AAD in NS\_Reveal decreases to 0.14, which remains significantly higher than that in NS\_RevealMore ($p$=0.029). However, Appendix \autoref{fig:aad-robust} suggests that during the last 5 rounds, AAD levels between NS\_Reveal and NS\_RevealMore become nearly indistinguishable, a finding corroborated by complementary regression analyses reported in Appendix \autoref{tab:aad-regression-robust}. This finding suggests that the efficiency gain associated with providing the additional lottery information tends to diminish as participants gain experience. This is nonetheless consistent with our earlier discussion at the end of Subsection \ref{theory:neighbor}, which posits that increased market stability may reduce the necessity for full revelation of lottery information. Overall, the analysis of AAD offers initial empirical support for Hypothesis \ref{exp1:hypothesis:3}.

	\begin{table}[!htb]
		
		\caption{The frequencies of deviating from equilibrium strategies (\%) } 
		\label{tab:equilibrium}    
		\begin{subtable}[t]{\textwidth}
			\centering
			\caption{NoNS\_Cover and NS\_Cover}
			\begin{tabular}{cccccccc}
				\hline
				\multicolumn{2}{c}{NoNS\_Cover}  & \multicolumn{6}{c}{NS\_Cover} \\ \cmidrule(lr){1-2} \cmidrule(lr){3-8} 
				Type-$\bv$ & Type-$\bv'$         & $ s_1 $-neighbor  & $ s_2 $-neighbor & \multicolumn{2}{c}{$ s_3 $-neighbor} & \multicolumn{2}{c}{others} \\ \cmidrule(lr){5-6} \cmidrule(lr){7-8}
				&                     &                   &                  &  Type-$\bv$ & Type-$\bv'$ &  Type-$\bv$ & Type-$\bv'$ \\ \hline
				38.98     & 47.47               & 0                 & 8.75             & 67.31       & 69.64       & 53.37       & 51.30 \\ 
				\hline 
			\end{tabular}
		\end{subtable}
		\hfill
		
		\begin{subtable}[t]{\textwidth}
			\centering
			\caption{NoNS\_Reveal and NS\_Reveal}
			\begin{tabular}{lccccc}
				\hline
				\multirow{2}{*}{lottery info} & NoNS\_Reveal & \multicolumn{4}{c}{NS\_Reveal} \\ \cmidrule(lr){2-2} \cmidrule(lr){3-6} 
				& all students & $ s_1 $-neighbor & $ s_2 $-neighbor & $ s_3 $-neighbor & others \\ \hline
				own lottery = 1 & 1.25 & 2.38 & 0     & 3.85  & 6.82 \\ 
				own lottery = 2 & 6.25 & 0    & 22.22 & 5.56  & 26.92 \\ 
				own lottery = 3 & 5.00 & 0    & 5.26  & 46.43 & 26.32 \\ 
				own lottery = 4 & 2.50 & 0    & 0     & 23.08 & 36.59 \\ 
				own lottery = 5 & 7.50 & 0    & 0     & 6.67  & 95.12 \\ 
				own lottery = 6 & 4.38 & 0    & 0     & 3.13  & 18.92 \\ \hline
			\end{tabular}
		\end{subtable}
		\hfill
		
		\begin{subtable}[t]{\textwidth}
			\centering
			\caption{NS\_RevealMore}
			\begin{tabular}{lcccc}
				\hline 
				lottery info & $ s_1 $-neighbor & $ s_2 $-neighbor & $ s_3 $-neighbor & others \\ \hline
				own lottery = 1 & 0 & 25$^a$& 3.85  & 3.41  \\ 
				own lottery = 2 & 0 & 16.67 & 0     & 14.10   \\ 
				own lottery = 3 & 0 & 0     & 14.29 & 10.53  \\ 
				own lottery = 4 & 0 & 0     & 7.69  & 21.95  \\ 
				own lottery = 5 & 0 & 0     & 6.67  & 9.76   \\ 
				own lottery = 6 & 0 & 0     & 3.13  & 8.11   \\ \hline
			\end{tabular}
		\end{subtable}
		\par
		\vspace{0.3cm}  {\footnotesize \textit{Notes:} $^a$ indicates that there is only 4 observations in that cell. }
	\end{table}

	Next, we examine individual strategies more closely to learn how the Reveal and RevealMore policies help improve matching market efficiency. First, we compute the frequency of deviating from a specific equilibrium strategy contingent upon students' roles and utility types. The structure of \autoref{tab:equilibrium} is similar to that of \autoref{tab:equilibrium-prediction}, with the key distinction that each equilibrium strategy is replaced by the observed frequency of deviating from that strategy. An exception is made for NS\_RevealMore in panel (c), in which we compute the deviation frequency conditional on students' own lottery number to facilitate more direct comparisons with NS\_Reveal. Later, we provide more detailed analyses on decisions that depend on both students' own lottery number and the lottery numbers of neighborhood students.
	
	\autoref{tab:equilibrium} shows that, in the absence of neighborhood schools, the Reveal policy significantly increases the likelihood that students with various lottery numbers adhere to equilibrium strategies. When neighborhood schools are available, the Reveal policy mainly benefits $s_3$-neighbor and no-neighborhood students by increasing their conformity to equilibrium strategies.\footnote{The Reveal policy also slightly reduces the overall frequency of deviation for $s_2$-neighbor students from 8.75\% to 6.25\%. However, $s_2$-neighbor students with lottery number 2 tend to make more mistakes than those with other lottery numbers, occasionally reporting $s_1$ despite an equilibrium strategy to always report $s_2$. } For example, $s_3$-neighbor students deviate only 15\% of the time, which is substantially lower than 68.13\% in NS\_Cover.\footnote{In previous literature, this behavioral pattern of $s_3$-neighbor students observed in NS\_Cover is referred to as neighborhood school bias \citep{calsamiglia2010constrained}. However, given that theoretical predictions vary across policies, we refrain from asserting that the Reveal policy directly mitigates this bias here. } Notably, the efficacy of the Reveal policy is not uniform across lottery numbers; students receiving either the best or worst lottery numbers deviate much less frequently than those with intermediate lottery numbers. As previously noted, the equilibrium strategy for no-neighborhood schools with lottery number 5 is unintuitive, causing almost all these students' decisions to deviate from the equilibrium strategy. However, even after excluding these observations, the tendency of $s_3$-neighbor and no-neighborhood students with intermediate lottery numbers making more mistakes remains persistent. To further investigate the sources of these mistakes, \autoref{tab:strategy} presents the frequencies of all possible submission strategies by students' roles and utility types. Compared to theoretical predictions summarized in \autoref{tab:strategy-prediction}, $s_3$-neighbor and no-neighborhood students tend to err on the side of being too conservative in their submission strategies under the Cover policy. Although the Reveal policy mitigates this tendency, particularly among type-$\bv$ students, it does not entirely eliminate it. This behavioral pattern suggests that $s_3$-neighbor and no-neighborhood students with intermediate lottery numbers face more complex strategic challenges, which the Reveal policy partially alleviates, relative to $s_1$-neighbor, $s_2$-neighbor and students with either the best or worst lottery numbers.
	
	Finally, panel (3) of \autoref{tab:equilibrium} shows that the RevealMore policy further enhances adherence to equilibrium strategies among $s_3$-neighbor and no-neighborhood students, notwithstanding the persistence of relatively higher deviation rates among those with intermediate lottery numbers. For example, $s_3$-neighbor students deviate only 6.25\% of the time; even those with lottery number 3 err in just 14.29\% of cases, substantially lower than 46.43\% in NS\_Reveal. Appendix \autoref{tab:equilibrium-revealmore} reports more detailed analyses, allowing decisions to depend on both own lottery number and lottery numbers of neighborhood students. We observe that except in a few instances with limited observations (i.e., no more than 6), the frequencies of deviation generally remain below 20\%. One notable exception occurs when no-neighborhood students with lottery number 4 observe that both $s_1$-neighbor's and $s_2$-neighbor's lottery numbers exceed 2. In this case, although they should have reported $s_3$, they mistakenly report $s_2$ in 35\% of the time. This discrepancy likely reflects the strategic complexity inherent in the situation: despite holding a better-than-average lottery number of 3, which intuitively suggests a favorable chance of admission to $s_2$, the equilibrium strategy indicates otherwise. 
	
	In sum, these detailed analyses of individual strategies provide evidence consistent with Hypothesis \ref{exp1:hypothesis:3}. Revealing more lottery information enables students to make more informed decisions that align more closely with equilibrium strategies.

	\begin{table}[!htb]
		\caption{The frequencies of all possible submission strategies (\%)}
		\label{tab:strategy}
		\begin{center}
			\begin{adjustbox}{width=1\textwidth}
				\begin{tabular}{lcccccccccc}
					\hline
					& \multicolumn{5}{c}{Type-$\bv$}                                      & \multicolumn{5}{c}{Type-$\bv'$}                                 \\
					\cmidrule(lr){2-6} \cmidrule(lr){7-11}
					Strategy	      & all     & $s_1$-neighbor & $s_2$-neighbor & $s_3$-neighbor & other  & all     & $s_1$-neighbor & $s_2$-neighbor & $s_3$-neighbor & other  \\
					\hline
					\multicolumn{11}{l}{\textbf{NoNS\_Cover}} \\
					$s_1$           & 61.02   & & & &                                                     & 39.87   & & & &  \\
					$s_2$           & 29.04   & & & &                                                     & 52.53   & & & &  \\
					$s_3$           & 9.94    & & & &                                                     & 7.59    & & & &  \\
					\multicolumn{11}{l}{\textbf{NoNS\_Reveal}} \\
					$s_1$           & 35.09   & & & &                                                     & 30.06   & & & &  \\
					$s_2$           & 31.83   & & & &                                                     & 39.56   & & & &  \\
					$s_3$           & 33.07   & & & &                                                     & 30.38   & & & &  \\
					\multicolumn{11}{l}{\textbf{NS\_Cover}} \\
					$s_1$           &     & 100            & 8.93           & 32.69          & 46.63      &     & 100           & 8.33            & 8.93           & 37.01 \\
					$s_2$           &     & 0              & 91.07          & 16.35          & 33.13      &     & 0             & 91.67           & 30.36          & 48.70 \\
					$s_3$           &     & 0              & 0              & 50.96          & 20.25      &     & 0             & 0               & 60.71          & 14.29 \\
					\multicolumn{11}{l}{\textbf{NS\_Reveal}} \\
					$s_1$           &     & 100            & 8.93           & 22.12          & 25.77      &     & 98.28         & 8.33            & 12.50          & 18.18 \\
					$s_2$           &     & 0              & 91.07          & 16.35          & 30.98      &     & 1.72          & 91.67           & 39.29          & 46.75 \\
					$s_3$           &     & 0              & 0              & 61.54          & 43.25      &     & 0             & 0               & 48.21          & 35.06 \\
					\multicolumn{11}{l}{\textbf{NS\_RevealMore}} \\
					$s_1$           &     & 100            & 10.71          & 20.19          & 25.77      &     & 100           & 6.25            & 21.43          & 20.13 \\
					$s_2$           &     & 0              & 89.29          & 24.04          & 30.06      &     & 0             & 93.75           & 32.14          & 38.31 \\
					$s_3$           &     & 0              & 0              & 55.77          & 44.17      &     & 0             & 0               & 46.43          & 41.56 \\
					\hline			
				\end{tabular}
			\end{adjustbox}
		\end{center}
	\end{table}

	\subsection{A Robustness Experiment in More Complex Environments}\label{sec:exp2}
	
	Our main experiment examines scenarios in which students are permitted to report only one school in their preference list. The matching market is relatively small and the cardinal preferences are somewhat arbitrary. However, this setup is essential for making precise theoretical predictions, but perhaps at the cost of realism. To partially address this limitation, the robustness experiment considers a larger matching market with neighborhood schools (16 students and 4 schools), allowing students to report up to two schools. All students' ordinal preferences are identical, but their cardinal preferences incorporate multiple random components. The principal findings are consistent across both experiments: the Reveal policy improves the overall match rate and market efficiency compared to the Cover policy, and students generally make more informed decisions based on observed lottery numbers. A detailed report on the robustness experiment is provided in Online Appendix \ref{appendix:exp2}.

	\section{Conclusion}
	Lotteries are widely used in school choice systems to break priority ties. This paper advocates for a policy that reveals lottery results to students before preference submission, contrasting it with the prevailing practice of withholding such information. Our theoretical analysis shows that when students face constraints on the length of their preference lists, which is a common feature in practice, the reveal policy improves students' decision-making, enabling them to coordinate their choices more effectively. This, in turn, enhances matching outcomes and increases overall welfare. Complementary lab experiments support these findings, showing that the reveal policy leads to higher match rates, higher student payoffs, and better alignment between reported preferences and achievable options.
	
	The recent adoption of the reveal policy by New York City indicates a progressive shift towards more transparent and equitable admission processes. Future research should explore the long-term implications of such policy changes, particularly as more data become available to assess their impact on school choice dynamics. By providing students with vital information, the reveal policy not only promotes informed decision-making but also fosters fairness, ultimately contributing to more efficient school allocation. This study highlights the essential role of transparency in shaping effective admission mechanisms in educational settings.
	
	\hide{
		\section*{Declaration of generative AI and AI-assisted technologies in the writing process}
		
		During the preparation of this work the author(s) used ChatGPT in order to proofread the paper. After using this tool/service, the author(s) reviewed and edited the content as needed and take(s) full responsibility for the content of the published article.
	}
	

	\clearpage

	\appendix
	
	

	
	
	
	\section*{Appendix: Additional Figures and Tables}\label{appendix:additionalresults}
	
	\setcounter{figure}{0}
	\setcounter{table}{0}
	\renewcommand\thetable{A\arabic{table}}
	\renewcommand\thefigure{A\arabic{figure}}
	
	\begin{figure}[!htb]
		\begin{center}
			\caption{Match rate over round}
			\label{fig:match-rate}
			\includegraphics[width=\linewidth]{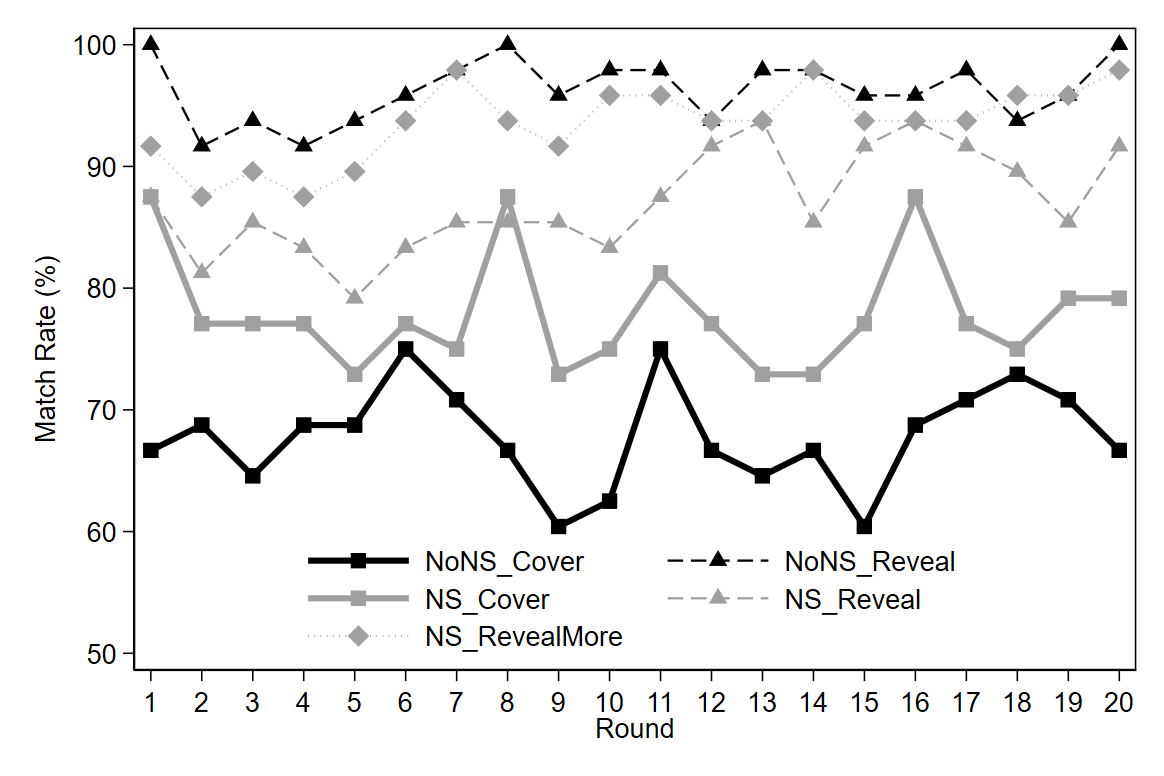}	
		\end{center}
	\end{figure}

	\begin{figure}[!htp]
		\caption{Average absolute deviation (excluding no-neighborhood students with lottery number 5)}
		\begin{center}
			\begin{subfigure}[b]{0.47\textwidth}
				\centering
				\caption{Average}
				\label{fig:deviation-robust}
				\includegraphics[width=\linewidth]{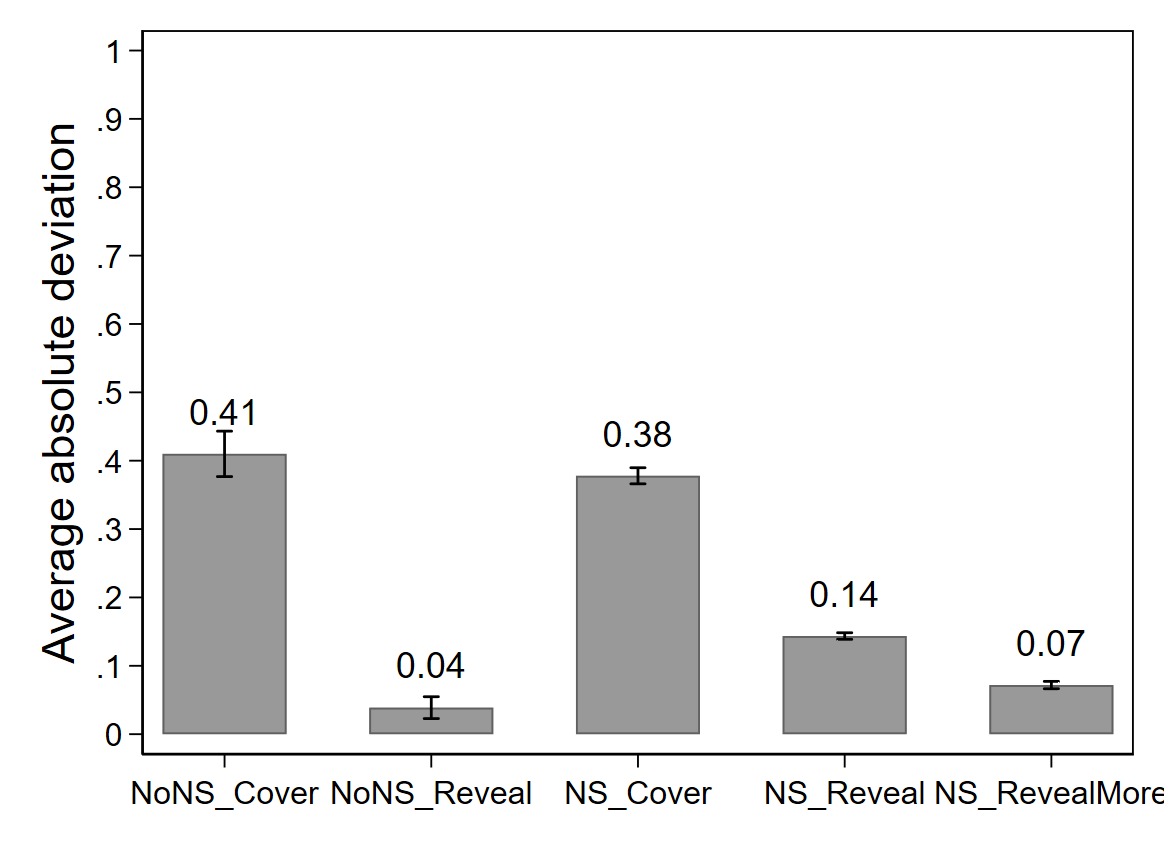}
			\end{subfigure}
			\begin{subfigure}[b]{0.47\textwidth}
				\centering
				\caption{Across rounds}
				\label{fig:deviation-round-robust}
				\includegraphics[width=\linewidth]{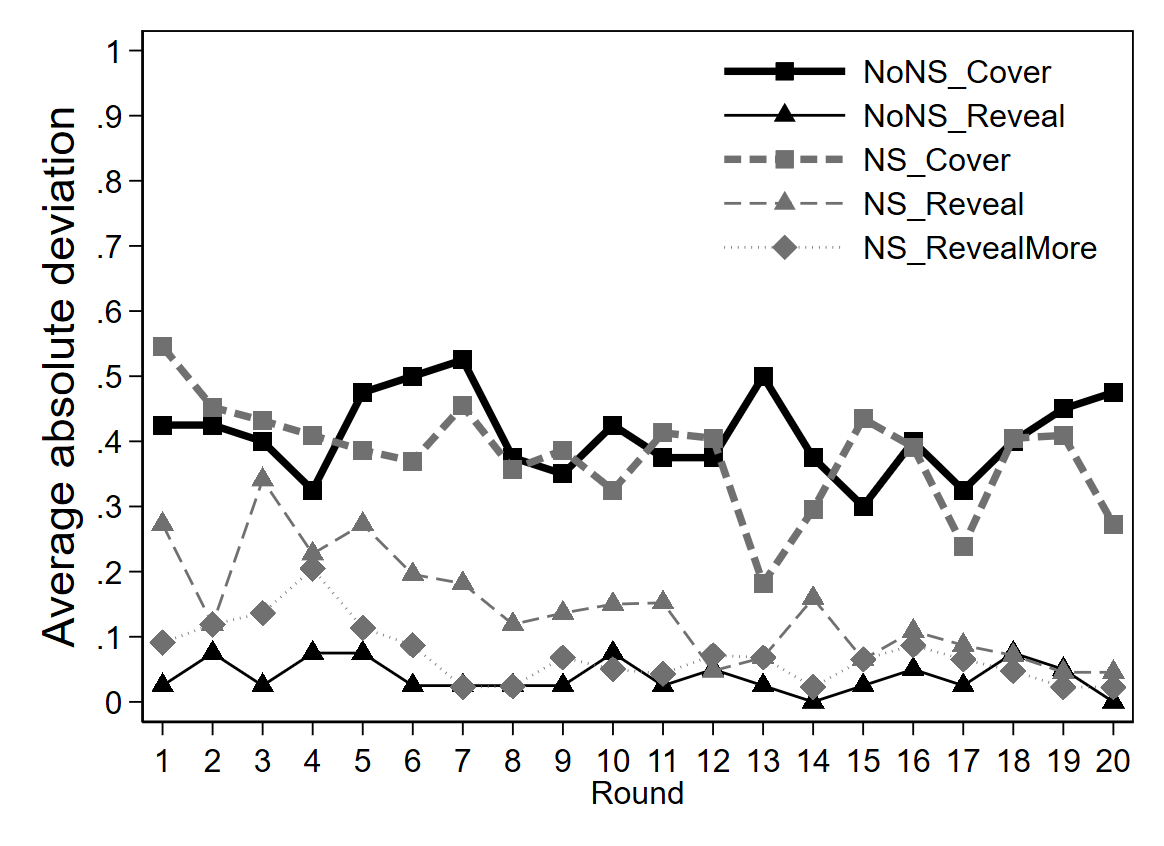}
			\end{subfigure}
		\end{center}
		\label{fig:aad-robust}
		\par
		\vspace{-0.5cm}  {\footnotesize \textit{Notes:} Standard errors clustered at the matching group level are indicated by bars. }
	\end{figure}

	\begin{table}[!htb]
		\caption{Random effects probit regressions for decisions deviating from equilibrium strategies}
		\label{tab:aad-regression}
		\par
		\begin{center}
			\begin{tabular}{lcc}
				\hline
				& All rounds    & Last 5 rounds \\ \hline
				NoNS\_Reveal        &      -0.381***&      -0.364***\\
				&     (0.035)   &     (0.030)   \\
				NS\_Cover           &      -0.023   &      -0.050   \\
				&     (0.036)   &     (0.034)   \\
				NS\_Reveal          &      -0.204***&      -0.270***\\
				&     (0.035)   &     (0.031)   \\
				NS\_RevealMore      &      -0.349***&      -0.348***\\
				&     (0.034)   &     (0.030)   \\
				Round               &      -0.006***&      -0.003   \\
				&     (0.001)   &     (0.009)   \\
				&               &               \\
				Clusters            & 20            & 20            \\
				N                   & 4800          & 1200          \\ \hline
				H0: NS\_Reveal = NS\_RevealMore & $p$<0.001  & $p$<0.001 \\ \hline
			\end{tabular}
		\end{center}
		\par
		{\footnotesize \textit{Notes:} Standard errors clustered at the matching group level are in parentheses. NoNS\_Cover serves as the benchmark. $^{*}$ $p < 0.1$, $^{**}$ $p < 0.05$, $^{***}$ $p < 0.01$. }
	\end{table}

	\begin{table}[!htb]
		\caption{Random effects probit regressions for decisions deviating from equilibrium strategies (excluding no-neighborhood students with lottery number 5)}
		\label{tab:aad-regression-robust}
		\par
		\begin{center}
			\begin{tabular}{lcc}
				\hline
				& All rounds    & Last 5 rounds \\ \hline
				NoNS\_Reveal        &      -0.382***&      -0.376***\\
				&     (0.035)   &     (0.030)   \\
				NS\_Cover           &      -0.033   &      -0.069*  \\
				&     (0.037)   &     (0.037)   \\
				NS\_Reveal          &      -0.273***&      -0.344***\\
				&     (0.035)   &     (0.030)   \\
				NS\_RevealMore      &      -0.348***&      -0.366***\\
				&     (0.034)   &     (0.030)   \\
				Round               &      -0.006***&      -0.007   \\
				&     (0.001)   &     (0.010)   \\
				&               &               \\
				Clusters            & 20            & 20            \\
				N                   &    4234       &    1066          \\ \hline
				H0: NS\_Reveal = NS\_RevealMore & $p$<0.001  & $p$=0.104 \\ \hline                    
			\end{tabular}
		\end{center}
		\par
		{\footnotesize \textit{Notes:} Standard errors clustered at the matching group level are in parentheses. NoNS\_Cover serves as the benchmark. $^{*}$ $p < 0.1$, $^{**}$ $p < 0.05$, $^{***}$ $p < 0.01$. }
	\end{table}

	\begin{table}[!htb]
		\centering
		\caption{The frequencies of deviating from equilibrium strategies (\%) in NS\_RevealMore}
		\label{tab:equilibrium-revealmore} 
		\begin{tabular}{lcccc}
			\hline 
			lottery info & $ s_1 $-neighbor & $ s_2 $-neighbor & $ s_3 $-neighbor & others \\ \hline
			own lottery = 1    & 0                    & 25$^a$                  & 3.85                   & 3.41 \\ 
			own lottery = 2 \& & \multirow{2}{*}{N/A} & \multirow{2}{*}{12.50}  & \multirow{2}{*}{0}     & \multirow{2}{*}{19.23} \\
			\hspace{10pt} $ s_1 $-neighbor's lottery = 1 \\ 
			own lottery = 2 \& & \multirow{2}{*}{0}   & \multirow{2}{*}{17.86}  & \multirow{2}{*}{0}     & \multirow{2}{*}{11.5} \\
			\hspace{10pt} $ s_1 $-neighbor's lottery > 1 \\ 
			own lottery = 3 \& & \multirow{2}{*}{N/A} & \multirow{2}{*}{0}      & \multirow{2}{*}{11.11} & \multirow{2}{*}{0} \\
			\hspace{10pt} $ s_1 $-neighbor's lottery $\le$ 2 \\ 
			own lottery = 3 \& & \multirow{3}{*}{0$^a$}& \multirow{3}{*}{N/A}   & \multirow{3}{*}{0$^a$}     & \multirow{3}{*}{4.55} \\ 
			\hspace{10pt} $ s_1 $-neighbor's lottery > 2 \&   \\
			\hspace{10pt} $ s_2 $-neighbor's lottery $\le 2$ \\ 
			own lottery = 3 \& & \multirow{3}{*}{0}   & \multirow{3}{*}{0}      & \multirow{3}{*}{50$^a$}& \multirow{3}{*}{35.00} \\ 
			\hspace{10pt} other cases  \\
			own lottery = 4 \& & \multirow{3}{*}{N/A} & \multirow{3}{*}{N/A}    & \multirow{3}{*}{16.67$^a$} & \multirow{3}{*}{16.67} \\ 
			\hspace{10pt} $ s_1 $-neighbor's lottery $\le 3$ \&   \\
			\hspace{10pt} $ s_2 $-neighbor's lottery $\le 3$ \\ 
			own lottery = 4 \& & \multirow{3}{*}{0}   & \multirow{3}{*}{0}      & \multirow{3}{*}{5.00}  & \multirow{3}{*}{24.14} \\ 
			\hspace{10pt} other cases   \\
			own lottery = 5    & 0                    & 0                       & 6.67                   & 9.76 \\ 
			own lottery = 6    & 0                    & 0                       & 3.13                   & 8.11 \\ \hline
		\end{tabular}
		\par
		\vspace{0.3cm}  {\footnotesize \textit{Notes:} $^a$ indicates that there are no more than 6 observations in that cell. N/A indicates no observation in that cell. }
	\end{table}
	
	\clearpage
	
	\setlength{\bibsep}{0pt plus 0.3ex}
	\bibliographystyle{aea}
	\bibliography{reference}
	
	\clearpage
	
	\section*{Online Appendix}
	
	\setcounter{page}{1}
	
	\section{Experimental Instructions}\label{appendix:instructions}

\textit{The following instructions are translated from the original instructions in Chinese. We provide the instructions for three treatments with neighborhood schools. The texts which differ between the revealing and covering treatments are highlighted in both italicized and bold face.}

\textbf{General Instructions}

You are participating in a decision-making experiment. All participants receive the same experimental instructions, so please read them carefully. During the experiment, no communication is allowed between participants, so please put your phone on silent mode. If you have any questions, feel free to raise your hand, and the experiment staff will come to assist you. You have earned 15 RMB for showing up on time. You can earn additional experimental rewards based on the decisions you make during the experiment. After the experiment concludes, the total points you earn will be converted into RMB at a rate of 1 points = 1 yuan. The final payment will be paid via bank transfer within 2-3 working days. The decisions made by participants in the experiment are completely anonymous, meaning your name will be strictly confidential in the study, and other participants will not know the total experimental rewards you have received today.

In today's experiment, we will simulate the process of students' admission to colleges. You and other participants will play the roles of students, and each of you will submit a preference list of schools to the admission system. The system will then assign an admission result to each student. Below are the descriptions of the experimental steps, payoff rules, and admission rules.

\textbf{Experimental Steps:}

\begin{itemize}
    \item The experiment consists of 20 rounds of decisions. Before the experiment starts, you will be randomly assigned to a 6-person group along with other participants, and each person in the group will play the role of a student. Each group will face 3 different schools, labeled as A, B, and C. Each school has 2 admission slots, and each slot can admit one student. The admissions for the 3 schools will be simulated by the computer.
    \item In each round of the experiment, participants will be randomly assigned to new groups. Therefore, it is unlikely that you will be grouped with the same five participants again.
    \item In each round, you will see the following payoff table, which shows the rewards you can get when being admitted to each school, representing your preferences for different schools. Your payoff depends on the school that admits you. For example, if your payoff table is as follows,
    \begin{table}[!h]
        \centering
        \begin{tabular}{|c|c|c|c|c|c|}
    	\hline
    	Admitted School & A & B & C & Not Admitted \\
    	\hline
    	Payoff (Points) & X & Y & Z & 0 \\
    	\hline
	\end{tabular}
    \end{table}
	
    Then:
	\begin{itemize}
		\item If School A admits you in a certain round, your payoff for that round will be X points.
    	\item If School B admits you in a certain round, your payoff for that round will be Y points. 
    	\item If School C admits you in a certain round, your payoff for that round will be Z points.
    	\item If no school admits you in a certain round, your payoff for that round will be 0 points. 
	\end{itemize}	 
    \item In each round, the payoff table for each student is randomly generated by the computer based on the following rules: 

    With a probability of two-thirds, the payoff table is:
    \begin{table}[!h]
        \centering
        \begin{tabular}{|c|c|c|c|c|c|}
    	\hline
    	Admitted School & A & B & C & Not Admitted \\
    	\hline
    	Payoff (Points) & 90 & 40 & 20 & 0 \\
    	\hline
	\end{tabular}
    \end{table}

    With another probability of one-third, the payoff table is:
    \begin{table}[!h]
        \centering
        \begin{tabular}{|c|c|c|c|c|c|}
    	\hline
    	Admitted School & A & B & C & Not Admitted \\
    	\hline
    	Payoff (Points) & 70 & 60 & 20 & 0 \\
    	\hline
	\end{tabular}
    \end{table}
    
    Therefore, in both payoff tables, students always attain the highest reward from being admitted to School A, followed by School B, with School C trailing last.
    
    \item In each round, every student's payoff table is independently and randomly generated in the manner described above. Regardless of the specific payoff table you receive, the probability that any other student within the same group possesses the first type of payoff table is two-thirds, while the probability of their having the second type is one-third.
    \item In each round, each student needs to submit a preference list. You can only submit one school out of the three schools A, B, or C in your preference list. 
    \item In each round, each student will be randomly assigned an ID within the group. In different rounds, each student's ID will be regenerated. Based on the student's ID, three students within the group will be viewed as residing in certain school districts, whereas the other three students do not belong to any district. Specifically:
    \begin{itemize}
    	\item Student with ID 1 resides in the district of School A.
    	\item Student with ID 2 resides in the district of School B. 
    	\item Student with ID 3 resides in the district of School C. 
    \end{itemize}
\end{itemize}

\textbf{Admission Rules:}

\begin{itemize}
    \item Lottery: In each round, the students will be assigned random numbers from 1 to 6, with no two students receiving the same number.
    
    Note: The lottery number is independent of the student's ID. In different rounds of the experiment, each student's ID and lottery number will be regenerated.
    
	\item Priority Ranking: During the admission process, each school sorts the students based on the ``within-district first, outside-district later'' rule, and then makes the admissions. Specifically, each school divides all students into two categories: 
	\begin{itemize}
		\item High Priority: Students residing in the school district.
    	\item Low Priority: Students residing outside the school district. 
	\end{itemize}

    Therefore, for each school, the high-priority group comprises one student, while the low-priority group consists of five students. Within the low-priority group, the five students are further ranked based on their assigned lottery numbers by the computer, with students receiving smaller numbers attaining higher ranks.

    \item Preference Submission: \textit{\textbf{[NS\_Cover: In each round, every student submits a single school preference. Subsequently, a computerized lottery is conducted. Hence, prior to submitting their preferences, no student is aware of their lottery number.]}} \textit{\textbf{[NS\_Reveal: In each round, the computer first conducts a lottery. Subsequently, each student is informed of their own lottery number. Finally, every student submits a preference for a single school.]}} \textit{\textbf{[NS\_RevealMore: In each round, the computer first conducts a lottery. Subsequently, each student is informed of their own lottery number, as well as the lottery numbers of the students residing within each school's district. Finally, every student submits a preference for a single school.]}}
    
    \item Admission Mechanism: In each round, once every student in each group has submitted their preferences, their applications are forwarded to the schools of their choice. Upon receiving these applications, each school ranks all applicants according to the above rule. Since each school can admit a maximum of two students, if the number of applications exceeds two, only the top two ranked students are admitted, while the remainder are rejected. The final admission outcome for all rejected students is non-admission.

\end{itemize}

\textbf{An Example:}

Let's use an example to further explain the admission rules. 

\textbf{Lottery Results:} Let's assume the lottery results are as shown below (lower numbers indicate higher rankings): \\
\begin{table}[!h]
    \centering
    \begin{tabular}{|c|c|c|c|c|c|c|}
        \hline
        Lottery Number & 1 & 2 & 3 & 4 & 5 & 6 \\
        \hline
        Student ID & 3 & 6 & 4 & 5 & 2 & 1 \\
        \hline
    \end{tabular}
\end{table}

\textbf{Priority Ranking:} According to the district rules, each school's priority ranking for all students is as follows (the numbers in the table represent student IDs): 
\begin{table}[!h]
    \centering
    \begin{tabular}{|c|c|c|}
        \hline
        School & High Priority (District) & Low Priority (Non-District) \\
        \hline
        A & 1 & 2, 3, 4, 5, 6 \\
        \hline
        B & 2 & 1, 3, 4, 5, 6 \\
        \hline
        C & 3 & 1, 2, 4, 5, 6 \\
        \hline
    \end{tabular}
\end{table}

\textbf{Final Sorting After Lottery:} Based on the district rules and lottery results, each school's sorting of students during admissions is as follows: \\
\begin{table}[!h]
    \centering
    \begin{tabular}{|c|c|c|c|c|c|c|}
        \hline
        School & Sort 1 (District) & Sort 2 & Sort 3 & Sort 4 & Sort 5 & Sort 6 \\
        \hline
        A & 1 & 3 & 6 & 4 & 5 & 2 \\
        \hline
        B & 2 & 3 & 6 & 4 & 5 & 1 \\
        \hline
        C & 3 & 6 & 4 & 5 & 2 & 1 \\
        \hline
    \end{tabular}
\end{table}

\textbf{Preference List:} \textit{\textbf{[NS\_RevealMore: In the experiment, each student is informed not only of their own lottery number but also of the numbers of all students within each school's district, providing the following lottery information:
\begin{table}[!h]
    \centering
    \begin{tabular}{|c|c|c|c|}
        \hline
        Student ID & 1 & 2 & 3  \\
        \hline
        District School & A & B & C  \\
        \hline
        Lottery Number  & 6 & 5 & 1  \\
        \hline
    \end{tabular}
\end{table}
]}}

Let's assume the students submit the following preference lists: \\
\begin{table}[!h]
    \centering
    \begin{tabular}{|c|c|c|c|c|c|c|}
        \hline
        Student ID & 1 & 2 & 3 & 4 & 5 & 6 \\
        \hline
        School & A & B & A & A & B & A \\
        \hline
    \end{tabular}
\end{table}

\textbf{Admissions Process:} Send each student's application to their ranked school. 

- School A admits students 1 and 3, rejects students 4 and 6. 

- School B admits students 2 and 5. 

\begin{table}[!h]
    \centering
    \begin{tabular}{cc|c|ccc}
        Application & & School & & Admitted & Rejected \\ \hline
        1,3,4,6 & $\rightarrow$ & A & $\rightarrow$ & 1,3 & 4,6 \\
        2,5 & $\rightarrow$ & B & $\rightarrow$ & 2,5 & \\
        & $\rightarrow$ & C & $\rightarrow$ & & \\
        \hline
    \end{tabular}
\end{table}

The final admission results are as follows: \\

\begin{table}[!h]
    \centering
    \begin{tabular}{|c|c|c|c|c|c|c|}
        \hline
        Student ID & 1 & 2 & 3 & 4 & 5 & 6 \\
        \hline
        Admitted School & A & B & A & Not Admitted & B & Not Admitted \\
        \hline
    \end{tabular}
\end{table}

\textbf{Payoff Rules:}
\begin{itemize}
    \item At the end of each round of the experiment, each student will be informed of whether they have been admitted, the school they have been admitted to, and their payoff. Note that each student's admission result in each round is independent of their admission results in previous rounds.
    \item After all 20 rounds of the experiment, we will randomly select one round of admission results as your final payoff. Additionally, you will receive a participation fee of 15 RMB. Finally, you can also earn additional income from the post-experiment questionnaire. \\
\end{itemize}

	\clearpage
	
	\section{Robustness Experiment}\label{appendix:exp2}
	
	\setcounter{figure}{0}
	\setcounter{table}{0}
	\renewcommand\thetable{B\arabic{table}}
	\renewcommand\thefigure{B\arabic{figure}}

\subsection{Experimental Design}

\paragraph{Environment} We design a matching market involving 16 students (ID1 to ID16) and 4 schools (A, B, C, D). Each school can admit 4 students. Students are played by experimental participants, while schools are simulated by the computer.

\begin{itemize}
	\item Preferences: Students have identical ordinal preferences $ A \succ B \succ C \succ D $, but they may differ in their assigned cardinal utilities (or payoffs). Specifically, their payoffs (in integer) from attending each school are randomly drawn from uniform distributions as follows: 
	$ u_A \sim U \{81, 100\} $,  $ u_B \sim U \{61, 80\} $, $ u_C \sim U \{41, 60\} $, $ u_D \sim U \{21, 40\} $.

    \item Neighborhoods: Two students reside in the neighborhoods of schools B, C and D, respectively; however, no student resides in the neighborhood of school A. The rationale for this feature is that any students living in the neighborhood of school A would surely report school A and gain admission. Our design effectively eliminates this trivial case.

	\item Priorities: All students are in priority ties for school A. For each of the other schools, students living in its neighborhood have higher priority than the others.
	
	\item Mechanism: A single lottery is drawn uniformly at random to resolve priority ties. Subsequently, DA is employed to generate the matching.

\end{itemize}

\paragraph{Treatments} We compare the two lottery policies in a between-subject design: one treatment implementing the cover policy and the other adopting the reveal policy. 
In the reveal treatment, the lottery is first drawn, followed by the announcement of the lottery number to each student prior to her submission of the ROL. In the cover treatment, the lottery is drawn after all students have submitted their ROL. 
Within each treatment, we also vary the length of the ROL, which can be either one or two, across two distinct blocks. Participants will initially make decisions in the block where the ROL length is one, followed by decisions in the block where the ROL length is two. Each block comprises 20 rounds of decisions.


\paragraph{Hypothesis} For the same ROL length, compared to the cover treatment,	the reveal treatment results in higher match rates, a lower likelihood of reporting neighborhood schools, reduced segregation levels, and greater overall welfare for students.

\paragraph{Experimental Procedure}
The experiment was conducted at the Nanjing Audit University Economics Experimental Lab with a total of 192 university students, using the software z-Tree \citep{Fischbacher2007}. We ran six sessions for each treatment. Each session consisted of 16 participants who interacted for a total of 40 rounds, divided into two blocks of 20 rounds each. Within each group, each participant was randomly assigned a student ID (ranging from 1 to 16), a lottery number (also ranging from 1 to 16), and payoffs from attending each of the four schools; all of these values were distinct in each round. Specifically, we generated a sequence of the 6-tuple of ID, lottery number and four payoffs at the group-round level. We applied it to each session to alleviate the concern that any treatment difference is simply due to the different random numbers. After every round, all participants received feedback about whether and which school they were admitted to and their round payoff. At the end of the session, one round from each block was privately and randomly chosen for each participant as her payoff for that block, and the participant's total payoff was the sum of the payoffs from the two blocks. The experimental instructions are reproduced in the online appendix.

Upon arrival, participants were randomly seated at a partitioned computer terminal. The experimental instructions were given to participants in printed form and were also read aloud by the experimenter. The instructions contain a detailed example illustrating how the matching works. To further improve comprehension, we also orally explained the example to participants with the help of PowerPoint slides. Participants then completed a comprehension quiz before proceeding. At the end of the experiment, they completed a demographic questionnaire. A typical session lasted about 1.5 hours with average earnings of 70.4 RMB, including a show-up fee of 15 RMB.

\subsection{Experimental Results}

\begin{figure}[!htb]
    \begin{center}
	\caption{Match rate over round}
	\label{fig:match-rate-exp2}
	\includegraphics[width=\linewidth]{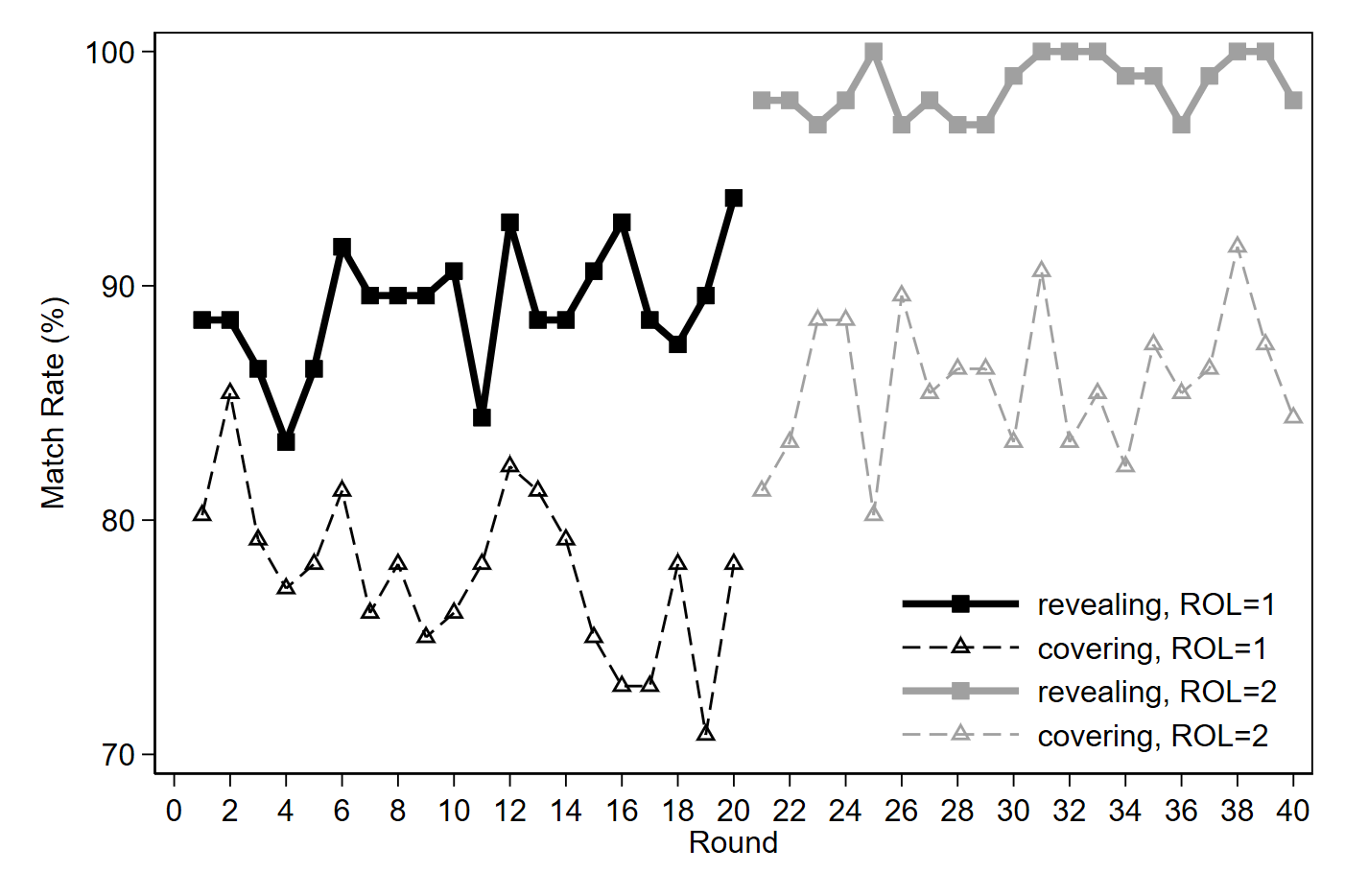}	
    \end{center}
\end{figure}

\paragraph{Match Rate} \autoref{fig:match-rate-exp2} shows the match rate across rounds under each condition. Consistent with our hypothesis, the reveal policy results in a significantly higher match rate than the cover policy, irrespective of the ROL length. When the ROL length is one, the average match rate is 89.1\% under the reveal policy, which is significantly higher than 77.8\% under the cover policy ($p$ = 0.004, Wilcoxon rank-sum test with each session considered as an independent observation). Similarly, when the ROL length is two, the rate is 98.5\% under the reveal policy compared to 85.9\% under the cover policy ($p$ = 0.003).

\paragraph{Neighborhood School Report Rate and Bias} The reveal policy is expected to assist students in making more rational decisions and thereby reduce their propensity to play safe by frequently reporting neighborhood schools. \autoref{table:district-rate} presents the neighborhood school report rate under each condition, focusing on students who have neighborhood schools. The table also presents a breakdown of this statistic by each of the four lottery categories. For instance, lottery (1, 4) refers to the four students whose lottery numbers are 1, 2, 3, and 4.

Consistent with our hypothesis, Panel A indicates that the reveal policy results in a significantly lower neighborhood school report rate compared to the cover policy regardless of the ROL length. The report rate across all lottery categories predictably varies under the reveal policy; students who have superior lottery numbers (lower number) are much less likely to report their neighborhood schools than those with inferior lottery numbers. Interestingly, when the ROL length is two, students are increasingly likely to report their neighborhood schools in the first rank as their lottery numbers become worse, even though this is a weakly dominated strategy (as these students could have weakly increased their chances of being admitted to a better school by ranking their neighborhood schools second). By contrast, under the cover policy, as students are not informed of their lottery numbers prior to making decisions, their report rates do not vary much across lottery categories. It is important to note when the ROL length is two, students rarely report their neighborhood schools in the first rank, which constitutes a rational decision. This finding suggests that the high frequency of reporting neighborhood schools in the first rank among students with inferior lottery numbers under the reveal policy is not due to participants' mistakes. Rather, it may simply reflect their beliefs that they could at best be admitted to their neighborhood schools, regardless of whether they rank it first or second.

\begin{table}[!htb]
    \begin{center}
    \caption{Neighborhood school report rate and bias}\label{table:district-rate}
    \begin{tabular}{lcccc}
        \hline
         & reveal & cover & p-value \\
        \hline
        \multicolumn{4}{l}{\textbf{Panel A: Neighborhood school report rate}} \\
        ROL=1 & 55.7\% & 67.5\% & 0.004 \\
        Lottery (1, 4) & 2.1\% & 65.1\% & 0.002 \\
        Lottery (5, 8) & 49.4\% & 64.7\% & 0.041 \\
        Lottery (9, 12) & 80.3\% & 70.2\% & 0.039 \\
        Lottery (12, 16) & 92.5\% & 69.5\% & 0.002 \\
        \addlinespace
        ROL=2 & 72.8\% (18.6\% + 54.2\%) & 87.8\% (3.9\% + 83.9\%) & 0.004 \\
        Lottery (1, 4) & 43.7\% (0.6\% + 43.1\%) & 89.1\% (5.7\% + 83.3\%) & 0.002 \\
        Lottery (5, 8) & 63.2\% (13.6\% + 49.6\%) & 86.0\% (3.5\% + 82.5\%) & 0.002 \\
        Lottery (9, 12) & 93.7\% (25.3\% + 68.4\%) & 90.8\% (2.3\% + 88.5\%) & 0.407 \\
        Lottery (12, 16) & 97.9\% (40.3\% + 57.6\%) & 85.4\% (4.2\% + 81.3\%) & 0.004 \\
        \addlinespace
        \multicolumn{4}{l}{\textbf{Panel B: Neighborhood school bias}} \\
        ROL=2 & 59.8\% (7.7\% + 52.1\%) & 81.9\% (1.0\% + 80.8\%) & 0.004\\
        Lottery (1, 4) & 9.3\% (0\% + 9.3\%) & 82.4\% (1.9\% + 80.6\%) & 0.002 \\
        Lottery (5, 8) & 46.2\% (0\% + 46.2\%) & 80.1\% (1.9\% + 78.2\%) & 0.002 \\
        Lottery (9, 12) & 90.8\% (10.8\% + 80.0\%) & 86.7\% (0\% + 86.7\%) & 0.407\\
        Lottery (12, 16) & 100\% (25.0\% + 75.0\%) & 78.1\% (0\% + 78.2\%) & 0.002 \\
        \hline
    \end{tabular}
    \end{center}
    {\footnotesize \textit{Notes:} Under ROL$=$2, the two percentages in the bracket represent the neighborhood school report rate in the first rank and in the second rank, respectively. \hide{The p-values are produced by the Wilcoxon rank-sum test which compares reveal and cover treatments using each session as an independent observation.} In Panel A, we focus on students who have neighborhood schools. In Panel B, we only focus on students whose neighborhood school is either C or D when the ROL length is two. }
\end{table}

When a student's neighborhood school is ranked high in payoffs, reporting that school does not necessarily imply neighborhood school bias. We define neighborhood school bias as the tendency to report neighborhood schools when the payoff rankings of these schools do not fall within the very top (when the ROL length is one) or the top two (when the ROL length is two). Under this definition, when the ROL length is one, the neighborhood school report rate is equivalent to the neighborhood school bias. However, when the ROL length is two, the scenario in which a student possesses a neighborhood school B and submits it in her ROL should not be interpreted as neighborhood school bias. Therefore, in Panel B, we only focus on students whose neighborhood schools are either C or D. Despite some level difference, we again find that the reveal policy results in a significantly lower neighborhood school bias compared to the cover policy, with other patterns qualitatively similar to those observed in Panel A.

\paragraph{Segregation} The reveal policy is expected to reduce not only the neighborhood school bias in students' ROL submissions but also the segregation level in the final admission outcome, measured as the proportion of students assigned to their neighborhood schools. \autoref{table:segregation} in Online Appendix \ref{appendix:additionalresults} indicates that the reveal policy results in a significantly lower segregation level compared to the cover policy, but this effect is significant only when the ROL length is one. 
When the ROL length is two, although the effect aligns with expectations, the reveal policy exhibits a weaker and statistically insignificant effect on the segregation level.  This is primarily driven by the luckiest students who could still be admitted to schools that are better than their neighborhood schools when they have more than one school to report. Thus, their neighborhood school bias has a much weaker impact on their final admission outcomes.

\begin{table}[!htb]
    \begin{center}
    \caption{Segregation (the proportion of subjects assigned to their neighborhood schools, only for those who have a neighborhood school)}\label{table:segregation}
    \begin{tabular}{lccc}
        \hline
         & reveal & cover & p-value \\
        \hline
        ROL=1 & 55.7\% & 67.5\% & 0.004 \\
        Lottery (1, 4) & 2.1\% & 65.1\% & 0.004 \\
        Lottery (5, 8) & 49.4\% & 64.7\% & 0.035\\
        Lottery (9, 12) & 80.3\% & 70.2\% & 0.027 \\
        Lottery (12, 16) & 92.5\% & 69.5\% & 0.004 \\
        \hline
        ROL=2 & 51.7\% & 52.8\% & 1.000 \\
        Lottery (1, 4) & 0.6\% & 5.7\% & 0.060 \\
        Lottery (5, 8) & 41.7\% & 43.4\% & 0.868 \\
        Lottery (9, 12) & 78.2\% & 85.1\% & 0.014 \\
        Lottery (12, 16) & 97.2\% & 85.4\% & 0.004 \\
        \hline
    \end{tabular}
    \end{center}
    {\footnotesize \textit{Notes:} To examine the segregation rate, we only focus on students who have neighborhood schools. The p-values are produced by the Wilcoxon rank-sum test which compares reveal and cover treatments using each session as an independent observation.}
\end{table}

\paragraph{Efficiency} \autoref{table:efficiency} reports the efficiency, measured by the student's average payoff, under each condition. Consistent with our hypothesis, the reveal policy results in significantly greater overall efficiency than the cover policy, irrespective of the ROL length. However, the benefits of the reveal policy are not uniformly distributed among students. While students with the best or worst lottery numbers tend to benefit from this policy, students with intermediate lottery numbers tend to suffer. The issue of unevenly distributed benefits is pronounced when the ROL length is one, but it is largely mitigated when the ROL length is two. Intuitively, when the possibility of mismatch is high (i.e., when the ROL length is short), prior knowledge of lottery numbers assists the most fortunate students in gaining admission to better schools while helping the least fortunate students avoid being left unmatched. However, this dynamic tends to disadvantage students in the middle, making them less likely to gain admission to top schools.


\begin{table}[!htb]
    \begin{center}
    \caption{Efficiency (mean payoff)}\label{table:efficiency}
    \begin{tabular}{lccc}
        \hline
         & reveal & cover & p-value   \\
        \hline
        ROL=1          & 54.6 & 51.8 & 0.007  \\
        Lottery (1, 4) & 87.3 & 68.0 & 0.004  \\
        Lottery (5, 8) & 54.5 & 64.2 & 0.004 \\
        Lottery (9, 12) & 42.6 & 46.8 & 0.016  \\
        Lottery (12, 16)& 33.9 & 28.1 & 0.004 \\
        \hline
        ROL=2          & 59.4 & 56.0 & 0.004 \\
        Lottery (1, 4) & 89.5 & 84.5 & 0.004  \\
        Lottery (5, 8) & 64.9 & 66.2 & 0.200  \\
        Lottery (9, 12)& 47.6 & 44.1 & 0.004  \\
        Lottery (12, 16) & 35.7 & 29.1 & 0.004  \\
        \hline
    \end{tabular}
    \end{center}
    {\footnotesize \textit{Notes:} The p-values are produced by the Wilcoxon rank-sum test which compares reveal and cover treatments using each session as an independent observation. }
\end{table}

\paragraph{ROL Submission Strategies} Following our review of the aggregate-level results, we turn to individual ROL strategies. \autoref{table:strategy} presents the frequency of each possible strategy under each condition. Under the reveal policy, we also calculate the frequency for each lottery category separately. When the ROL length is one, students tend to report better schools more frequently under the cover policy; in contrast, under the reveal policy, the frequency of reporting each school is relatively uniform. The reason is that students make more informed decisions under the reveal policy by primarily reporting schools where they have a realistic chance of admission: students whose lottery numbers range from 1 to 4 are most likely to report school A; students whose lottery numbers range from 5 to 8 are most likely to report school B; and so on. Similarly, when the ROL length is two, under the cover policy, students tend to report a pair of schools that more frequently include at least one top school (such as A-B and A-C) than two bottom schools (such as C-D); conversely, under the reveal policy, the reported pairs of schools are evenly distributed between top and bottom schools (such as A-B, B-C and C-D). Further, students tend to report schools to which they have realistic chances of admission based on their lottery numbers: students whose lottery numbers range from 1 to 4 are most likely to report the school pair A-B; students whose lottery numbers range from 5 to 8 are most likely to report the school pair B-C; students whose lottery numbers range from 9 to 16 are most likely to report the school pair C-D.\footnote{We also calculate the frequency of each possible strategy for students who have a specific neighborhood school (\autoref{table:strategy-district-B}, \autoref{table:strategy-district-C} and \autoref{table:strategy-district-D}) and for students who do not have any neighborhood school (\autoref{table:strategy-non-district}) separately. The results generally align with our expectations. For instance, students who have neighborhood school B only report school A or B when the ROL length is one and the school pairs A-B, B-C or B-D when the ROL length is two.
}

\begin{table}[!htb]
    \centering
    \caption{Proportion of each possible submission strategy (\%)}\label{table:strategy}
    \begin{adjustbox}{width=\textwidth}
    \begin{tabular}{lccccccc}
        \hline
        \multicolumn{2}{c}{Strategy} & \multicolumn{5}{c}{reveal}  & cover  \\
        \hline
        Rank 1 & Rank 2 & All & Lottery (1, 4) & Lottery (5, 8) & Lottery (9, 12) & Lottery (13, 16) & All \\
        \hline
        \multicolumn{2}{c}{ROL=1}  \\
        A & & 25.42 & \textbf{90.63} & 7.29 & 1.67 & 2.08 & 38.85 \\
        B & & 25.52 & 7.50 & \textbf{64.17} & 16.67 & 13.75 & 29.11 \\
        C & & 27.45 & 1.46 & 25.83 & \textbf{61.67} & 20.83 & 21.46 \\
        D & & 21.61 & 0.42 & 2.71 & 20.00 & \textbf{63.33} & 10.57 \\
        \hline
        \multicolumn{2}{c}{ROL=2}  \\
        A & B & 32.29 & \textbf{94.79} & \textbf{26.25} & 5.42 & 2.71 & 30.63 \\
        A & C & 1.61 & 1.67 & 1.25 & 2.92 & 0.63 & 29.43 \\
        A & D & 0.89 & 1.04 & 0 & 0.63 & 1.88 & 12.81 \\
        B & A & 0.78 & 0.83 & 1.25 & 0.42 & 0.63 & 1.56 \\
        B & C & 28.85 & 1.25 & \textbf{66.46} & \textbf{38.96} & 8.75 & 14.90 \\
        B & D & 2.81 & 0 & 2.50 & 2.92 & 5.83 & 7.45 \\
        C & A & 0.05 & 0 & 0 & 0.21 & 0 & 0.10 \\
        C & B & 0.52 & 0 & 0.21 & 1.67 & 0.21 & 0.05 \\
        C & D & 29.74 & 0.42 & 2.08 & \textbf{46.25} & \textbf{70.21} & 2.97 \\
        D & A & 0.21 & 0 & 0 & 0 & 0.83 & 0.05 \\
        D & B & 0 & 0 & 0 & 0 & 0 & 0 \\
        D & C & 2.24 & 0 & 0 & 0.63 & 8.33 & 0.05 \\
        \hline
    \end{tabular}
    \end{adjustbox}
\end{table}

\begin{table}[!htb]
    \centering
    \caption{Proportion of each possible submission strategy (\%) for students having neighborhood school B}\label{table:strategy-district-B}
    \begin{adjustbox}{width=\textwidth}
    \begin{tabular}{lccccccc}
        \hline
        \multicolumn{2}{c}{Strategy} & \multicolumn{5}{c}{reveal}  & cover  \\
        \hline
        Rank 1 & Rank 2 & All & Lottery (1, 4) & Lottery (5, 8) & Lottery (9, 12) & Lottery (13, 16) & All \\
        \hline
        \multicolumn{2}{c}{ROL=1} \\
        A & & 26.25 & 95.00 & 6.67 & 1.67 & 1.67 & 10.83 \\
        B & & 72.50 & 5.00 & 93.33 & 96.67 & 95.00 & 87.92 \\
        C & & 0.42 & 0 & 0 & 1.67 & 0 & 0.83 \\
        D & & 0.83 & 0 & 0 & 0 & 3.33 & 0.42 \\
        \hline
        \multicolumn{2}{c}{ROL=2} \\
        A & B & 58.33 & 98.48 & 56.94 & 42.59 & 22.92 & 90.00 \\
        A & C & 0 & 0 & 0 & 0 & 0 & 0 \\
        A & D & 0 & 0 & 0 & 0 & 0 & 0 \\
        B & A & 4.17 & 1.52 & 5.56 & 3.70 & 6.25 & 6.67 \\
        B & C & 30.00 & 0 & 37.50 & 53.70 & 33.33 & 2.50 \\
        B & D & 6.25 & 0 & 0 & 0 & 31.25 & 0.42 \\
        C & A & 0 & 0 & 0 & 0 & 0 & 0 \\
        C & B & 0 & 0 & 0 & 0 & 0 & 0 \\
        C & D & 1.25 & 0 & 0 & 0 & 6.25 & 0.42 \\
        D & A & 0 & 0 & 0 & 0 & 0 & 0 \\
        D & B & 0 & 0 & 0 & 0 & 0 & 0 \\
        D & C & 0 & 0 & 0 & 0 & 0 & 0 \\
        \hline
    \end{tabular}
    \end{adjustbox}
\end{table}

\begin{table}[!htb]
    \centering
    \caption{Proportion of each possible submission strategy (\%) for students having neighborhood school C}\label{table:strategy-district-C}
    \begin{adjustbox}{width=\textwidth}
    \begin{tabular}{lccccccc}
        \hline
        \multicolumn{2}{c}{Strategy} & \multicolumn{5}{c}{reveal}  & cover  \\
        \hline
        Rank 1 & Rank 2 & All & Lottery (1, 4) & Lottery (5, 8) & Lottery (9, 12) & Lottery (13, 16) & All \\
        \hline
        \multicolumn{2}{c}{ROL=1} \\
        A & & 22.08 & 91.67 & 12.96 & 0 & 3.70 & 15.83 \\
        B & & 15.00 & 8.33 & 51.85 & 4.76 & 0 & 12.92 \\
        C & & 61.67 & 0 & 35.19 & 94.05 & 92.59 & 71.25 \\
        D & & 1.25 & 0 & 0 & 1.19 & 3.70 & 0 \\
        \hline
        \multicolumn{2}{c}{ROL=2} \\
        A & B & 24.17 & 87.04 & 15.28 & 0 & 0 & 7.50 \\
        A & C & 7.50 & 9.26 & 8.33 & 7.58 & 4.17 & 73.75 \\
        A & D & 0 & 0 & 0 & 0 & 0 & 0 \\
        B & A & 0 & 0 & 0 & 0 & 0 & 0.42 \\
        B & C & 53.33 & 3.70 & 76.39 & 72.73 & 47.92 & 15.83 \\
        B & D & 0 & 0 & 0 & 0 & 0 & 0.83 \\
        C & A & 0 & 0 & 0 & 0 & 0 & 0.83 \\
        C & B & 1.67 & 0 & 0 & 4.55 & 2.08 & 0 \\
        C & D & 13.33 & 0 & 0 & 15.15 & 45.83 & 0.83 \\
        D & A & 0 & 0 & 0 & 0 & 0 & 0 \\
        D & B & 0 & 0 & 0 & 0 & 0 & 0 \\
        D & C & 0 & 0 & 0 & 0 & 0 & 0 \\
        \hline
    \end{tabular}
    \end{adjustbox}
\end{table}

\begin{table}[!htb]
    \centering
    \caption{Proportion of each possible submission strategy (\%) for students having neighborhood school D}\label{table:strategy-district-D}
    \begin{adjustbox}{width=\textwidth}
    \begin{tabular}{lccccccc}
        \hline
        \multicolumn{2}{c}{Strategy} & \multicolumn{5}{c}{reveal}  & cover  \\
        \hline
        Rank 1 & Rank 2 & All & Lottery (1, 4) & Lottery (5, 8) & Lottery (9, 12) & Lottery (13, 16) & All \\
        \hline
        \multicolumn{2}{c}{ROL=1} \\
        A & & 35.00 & 96.43 & 7.14 & 0 & 0 & 22.92 \\
        B & & 14.58 & 2.38 & 69.05 & 0 & 6.67 & 19.17 \\
        C & & 17.50 & 0 & 19.05 & 59.26 & 3.33 & 14.58 \\
        D & & 32.92 & 1.19 & 4.76 & 40.74 & 90.00 & 43.33 \\
        \hline
        \multicolumn{2}{c}{ROL=2} \\
        A & B & 28.33 & 90.74 & 21.43 & 1.85 & 0 & 13.75 \\
        A & C & 0 & 0 & 0 & 0 & 0 & 8.33 \\
        A & D & 1.67 & 5.56 & 0 & 0 & 2.08 & 47.92 \\
        B & A & 0 & 0 & 0 & 0 & 0 & 1.42 \\
        B & C & 27.50 & 3.70 & 65.48 & 16.67 & 0 & 5.00 \\
        B & D & 7.50 & 0 & 10.71 & 9.26 & 8.33 & 20.83 \\
        C & A & 0.42 & 0 & 0 & 1.85 & 0 & 0 \\
        C & B & 0 & 0 & 0 & 0 & 0 & 0 \\
        C & D & 34.17 & 0 & 2.38 & 70.37 & 87.50 & 3.33 \\
        D & A & 0 & 0 & 0 & 0 & 0 & 0.42 \\
        D & B & 0 & 0 & 0 & 0 & 0 & 0 \\
        D & C & 0.42 & 0 & 0 & 0 & 2.08 & 0 \\
        \hline
    \end{tabular}
    \end{adjustbox}
\end{table}

\begin{table}[!htb]
    \centering
    \caption{Proportion of each possible submission strategy (\%) for non-neighborhood students}\label{table:strategy-non-district}
    \begin{adjustbox}{width=\textwidth}
    \begin{tabular}{lccccccc}
        \hline
        \multicolumn{2}{c}{Strategy} & \multicolumn{5}{c}{reveal}  & cover  \\
        \hline
        Rank 1 & Rank 2 & All & Lottery (1, 4) & Lottery (5, 8) & Lottery (9, 12) & Lottery (13, 16) & All \\
        \hline
        \multicolumn{2}{c}{ROL=1} \\
        A & & 24.00 & 87.85 & 6.48 & 2.48 & 2.29 & 52.25 \\
        B & & 20.42 & 9.38 & 60.19 & 6.38 & 1.63 & 22.58 \\
        C & & 28.00 & 2.43 & 29.94 & 65.25 & 15.69 & 17.00 \\
        D & & 27.58 & 0.35 & 3.40 & 25.89 & 80.39 & 8.17 \\
        \hline
        \multicolumn{2}{c}{ROL=2} \\
        A & B & 29.50 & 96.08 & 22.22 & 0.65 & 0.60 & 26.75 \\
        A & C & 1.08 & 0.98 & 0 & 2.94 & 0.30 & 30.67 \\
        A & D & 1.08 & 0.65 & 0 & 0.98 & 2.38 & 10.92 \\
        B & A & 0.42 & 0.98 & 0.79 & 0 & 0 & 1.00 \\
        B & C & 24.00 & 0.65 & 72.22 & 33.01 & 0.89 & 19.17 \\
        B & D & 1.75 & 0 & 1.19 & 2.94 & 2.68 & 7.50 \\
        C & A & 0 & 0 & 0 & 0 & 0 & 0 \\
        C & B & 0.50 & 0 & 0.40 & 1.63 & 0 & 0.08 \\
        C & D & 37.83 & 0.65 & 3.17 & 56.86 & 80.36 & 3.83 \\
        D & A & 0.33 & 0 & 0 & 0 & 1.19 & 0 \\
        D & B & 0 & 0 & 0 & 0 & 0 & 0 \\
        D & C & 3.50 & 0 & 0 & 0.98 & 11.61 & 0.08 \\
        \hline
    \end{tabular}
    \end{adjustbox}
\end{table}

\end{document}